\documentclass[ aapm, mph, amsmath,amssymb,superscriptaddress,nofootinbib]{revtex4}
\usepackage{dcolumn}
\usepackage{bm}
\usepackage{color}
\usepackage{graphicx}
\usepackage{amsmath}
\usepackage{amssymb}
\usepackage{slashed}
\usepackage{tabularx}
\usepackage{wrapfig}
\usepackage{slashed}

\def\beq{\begin{equation}}
\def\eeq{\end{equation}}
\def\be{\begin{eqnarray}}
\def\ee{\end{eqnarray}}

\usepackage[mathlines]{lineno}

\begin{document}

\title{Exclusive final state hadron observables from neutrino-nucleus multi-nucleon knockout}

\author{J. E. Sobczyk}
\affiliation{Instituto de F\'\i sica Corpuscular (IFIC), Centro Mixto
CSIC-Universidad de Valencia, Institutos de Investigaci\'on de
Paterna, Apartado 22085, E-46071 Valencia, Spain}
\affiliation{Institut f\"ur Kernphysik and PRISMA$^+$ Cluster of Excellence, Johannes Gutenberg-Universit\"at Mainz, 55128 Mainz, Germany}

\author{J. Nieves}
\affiliation{Instituto de F\'\i sica Corpuscular (IFIC), Centro Mixto
CSIC-Universidad de Valencia, Institutos de Investigaci\'on de
Paterna, Apartado 22085, E-46071 Valencia, Spain}

\author{F. S\'{a}nchez}
 \affiliation{Universit\'{e} de Gen\`{e}ve - Facult\'{e} des Sciences, D\'{e}partement de Physique Nucl\'{e}aire et Corpusculaire (DPNC) \\
24, Quai Ernest-Ansermet, CH-1211 Gen\`{e}ve 4, Switzerland}

\date{\today}

\begin{abstract}

We present  results of an updated calculation of the 2p2h (two particle two hole) contribution to the neutrino-induced charge-current cross section. We provide also some exclusive observables, interesting from the point of view of experimental studies, e.g. distributions of momenta of the outgoing nucleons and of available energy, which we compare with the results obtained within the NEUT generator. We also compute, and separate from the total, the contributions of 3p3h mechanisms. 
Finally, we discuss the differences between the present results and previous implementations of the model in MC event-generators, done at the level of inclusive cross sections,  which  might significantly influence the experimental analyses, particularly in the cases where the hadronic observables are considered.

\end{abstract}
\maketitle

%
%
%
%
\section{Introduction}

The studies of neutrino-nucleus interactions are entering a new stage, motivated by long-baseline experimental programs, in which the statistical uncertainties will diminish and thus the nuclear effects -- contributing to the systematical error -- have to be kept well under control~\cite{Alvarez-Ruso:2017oui}. The incomplete theoretical knowledge of the neutrino-nucleus interactions influences various stages of experimental analysis.
For instance, for the future Hyper-Kamiokande water Cherenkov detector, the reconstruction of neutrino energy will be mainly based on the kinematical method, in which only the outgoing muon is observed and the reaction kinematics is assumed to be quasielastic (QE). However, the energy range of the neutrino flux produced in the J-PARC facility is such that other physical mechanisms give a non negligible contribution to the cross section. In particular,  multi-nucleon knockout processes (mainly driven by the excitation of two particle two hole, 2p2h, components in nuclei) should be taken into account. Since in the latter processes the interaction takes place on a pair of nucleons, the energy balance is different than in the QE case. Mismatching the signal coming from these two reaction mechanisms would lead to a bias in the energy reconstruction \cite{Nieves:2012yz,Alvarez-Ruso:2014bla}. It is therefore crucial to properly include the 2p2h channel into the Monte Carlo event generators.

Various theoretical groups have presented calculations for 2p2h contributions, providing mainly the results for inclusive cross sections.
There is an ongoing discussion on the treatment of this physical reaction. 
The topic, primarily explored theoretically few decades ago in the case of electron scattering, has found recently a new application for neutrino-nucleus interactions. The computation requires  multidimensional phase-space integration and the inclusion of $\Delta(1232)$ degrees of freedom. 
Most of the 2p2h approaches start from the in-medium  calculation of the meson exchange currents (MEC) between two nucleons, mediated via one pion. However, this treatment, in which a pion is the principal carrier of interaction, though adequate in the free space for moderate energies, requires substantial modifications inside of the nuclear medium,  where the situation becomes more complicated. The nuclear effects are modelled differently in various approaches.

The response functions for one- and two-body currents obtained from ab-initio calculations using the Green Function Monte Carlo approach have been presented in Ref.~\cite{Lovato:2013cua}. They may be treated as a benchmark for more approximated and phenomenological models, although there are two serious drawbacks from the point of view of experimental needs. First of all, the calculations are constrained to a limited phase space where the non-relativistic kinematics can be employed (nuclear correlations and electroweak currents are non-relativistic). Second, such approach provides only  inclusive cross section, without predicting the spectrum of outgoing nucleons.

Among other more phenomenological schemes, the model of Refs.~\cite{Martini:2009uj,Martini:2010ex} was the first one to be proposed for neutrino-nucleus 2p2h contributions. Conceptually it is closely related to the approach followed in the present work, with the computation of some many-body diagrams,  including  some $\Delta$h effects with the $\Delta$ self-energy in the nuclear medium obtained within the same formalism of Ref.~\cite{Oset:1987re}. The difference lies in the treatment of the non-resonant background contributions, which in these works were extrapolated from previous 2p2h calculations, either for  pion absorption at threshold~\cite{Shimizu:1980kb} or  for the ($e, e')$ inclusive reaction~\cite{Alberico:1983zg}.

Recently, an extension of electron-nucleus 2p2h calculation of Ref.~\cite{Benhar:2015ula} has been presented for the charge current (CC) and the neutral current (NC) cases \cite{Rocco:2018mwt}. In this approach, the correlations of the ground state are accounted for by means of hole spectral functions of both nucleons (the difference from the two-nucleon spectral function has been advocated to be small). The $\Delta$ exchange is parametrized as in Ref.~\cite{Hernandez:2007qq}, but only one-pion exchange between the nucleons is considered.
The approach of Ref.~\cite{Megias:2016fjk} is based on the same set of exchange currents as in  \cite{Rocco:2018mwt}, though, no correlations between initial nucleons are included (i.e. they are distributed according to the Fermi gas model).

Here, we follow the formalism derived in Ref.~\cite{Nieves:2011pp}, which together with the works Refs.~\cite{Martini:2009uj,Martini:2010ex}, provided the first sensible theoretical explanation~\cite{Nieves:2011yp,Nieves:2013fr} for the so-called MiniBooNE axial mass puzzle~\cite{AguilarArevalo:2010zc}. This theoretical calculation is widely used by the experimental community and it has been included into several MC event generators; still only at the level of inclusive cross sections, with the outgoing nucleons (produced at the primary vertex of interaction) being distributed isotropically according to the available phase space, and cascaded through the nucleus by means of a Monte Carlo algorithm. The final distributions, however, are only an approximation which does not fully take into account the internal dynamics of the process. 

We perform here a comparison between the results of exclusive final state hadron distributions, as they are obtained within the full model after undoing the outgoing nucleon phase-space integration,  and as they are implemented in the NEUT generator~\cite{Hayato:2009zz}. 
For the first time we also show the results of the model separately for the 2p2h and 3p3h contributions, which until now were treated together. This way we are able to make predictions of exclusive two-nucleon final states.
We also improve on the previous treatment of the in-medium interactions between nucleons and $\Delta$s, maintaining the energy-transfer dependence of $g^\prime_l$ and $g^\prime_t$, the longitudinal and transverse Landau-Migdal parameters of the effective nucleon-nucleon spin-isospin interaction~\cite{Oset:1987re}. The treatment of the $\Delta$ resonance  is also refined according to recent work (including  changes in the $\Delta$ propagator and in the dominant $C_5^A(q^2)$ electroweak form factor~\cite{Hernandez:2016yfb}). In addition,  the whole calculation is performed without the numerical approximations used in the previous work of Ref.~\cite{Nieves:2011pp}. Indeed, this work serves as a further validation of the previous calculation. Since there have been tensions observed between various theoretical approaches~\cite{Dolan:2019bxf}, we find both this confirmation and some further improvements of the model, an important step forward.

This work is organized in the following way. In Sec.~\ref{sec:formalism} we sketch the formalism, particularly concentrating on the description of the in-medium baryon-baryon effective interaction in Subsec.~\ref{sec:effective_int}, and the treatment of $\Delta$ self-energy and propagator in Subsecs.~\ref{sec:delta_treatment} and \ref{sec:refinements}. We also pay attention to the 3p3h mechanism in Subsec.~\ref{sec:3p3h}.
Next, in Sec.~\ref{sec:NEUT} we shortly describe how the model and final state interactions are implemented in NEUT event generator. The results are presented in Sec.~\ref{sec:results}. Firstly we focus on the inclusive cross sections, to understand the effect of the various refinements introduced in  the model. Afterwards we present the distributions of outgoing nucleons and an analysis of the available energy. The conclusions and outlook are presented in Sec.~\ref{sec:conclusions}.

%
%
%
%
\section{Formalism}
\label{sec:formalism}

The  multi-nucleon knockout formalism we employ is based on the approach introduced for neutrino-nucleus interaction for the first time in Ref.~\cite{Nieves:2011pp}. In the past, the model has been extensively used for electron-, photon-, pion-nucleus scattering, and proved to describe the available data with a good accuracy~\cite{Gil:1997bm, Carrasco:1989vq, Nieves:1991ye}.

\begin{figure}[h]
\centering
\includegraphics[scale=0.1]{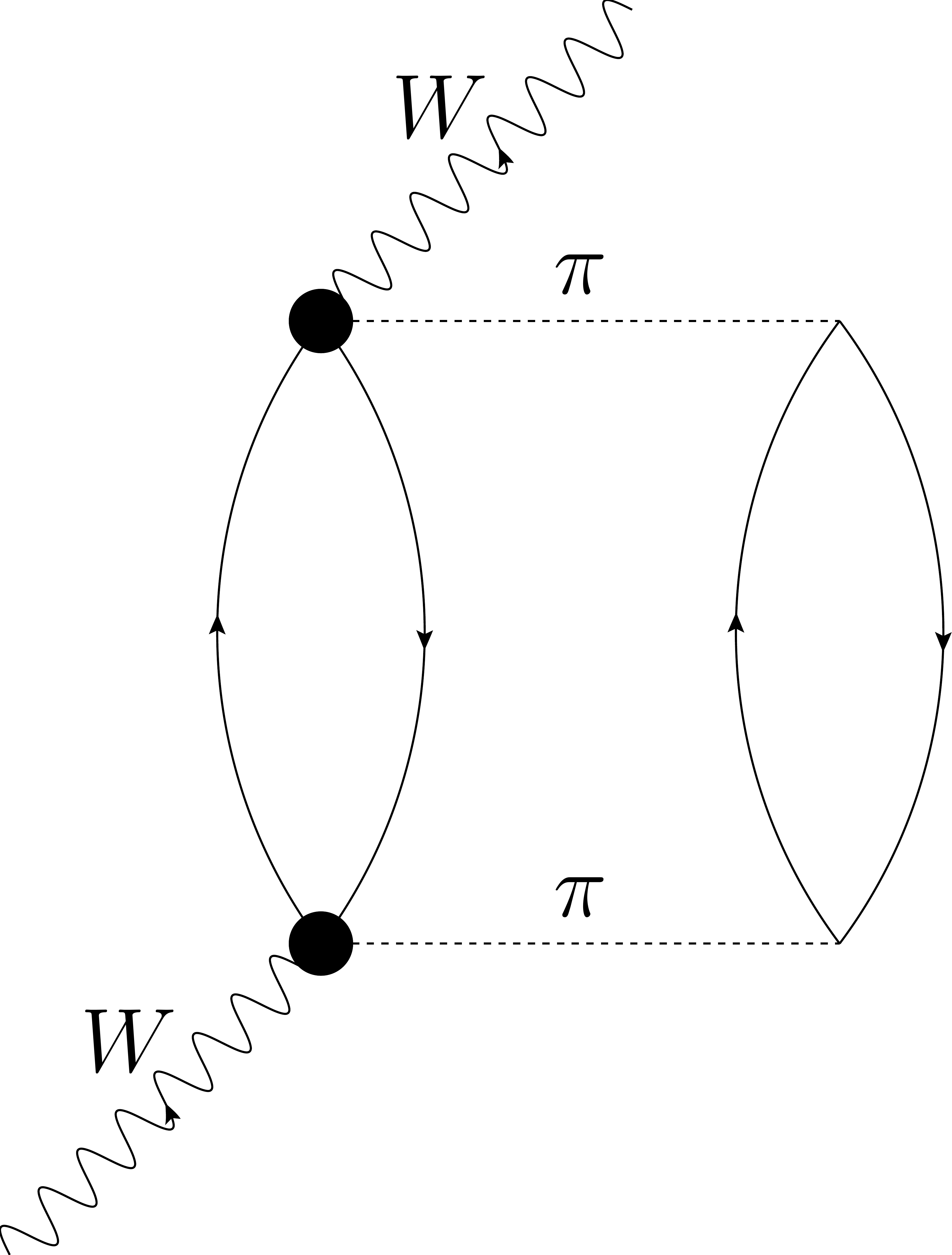}
\includegraphics[scale=0.1]{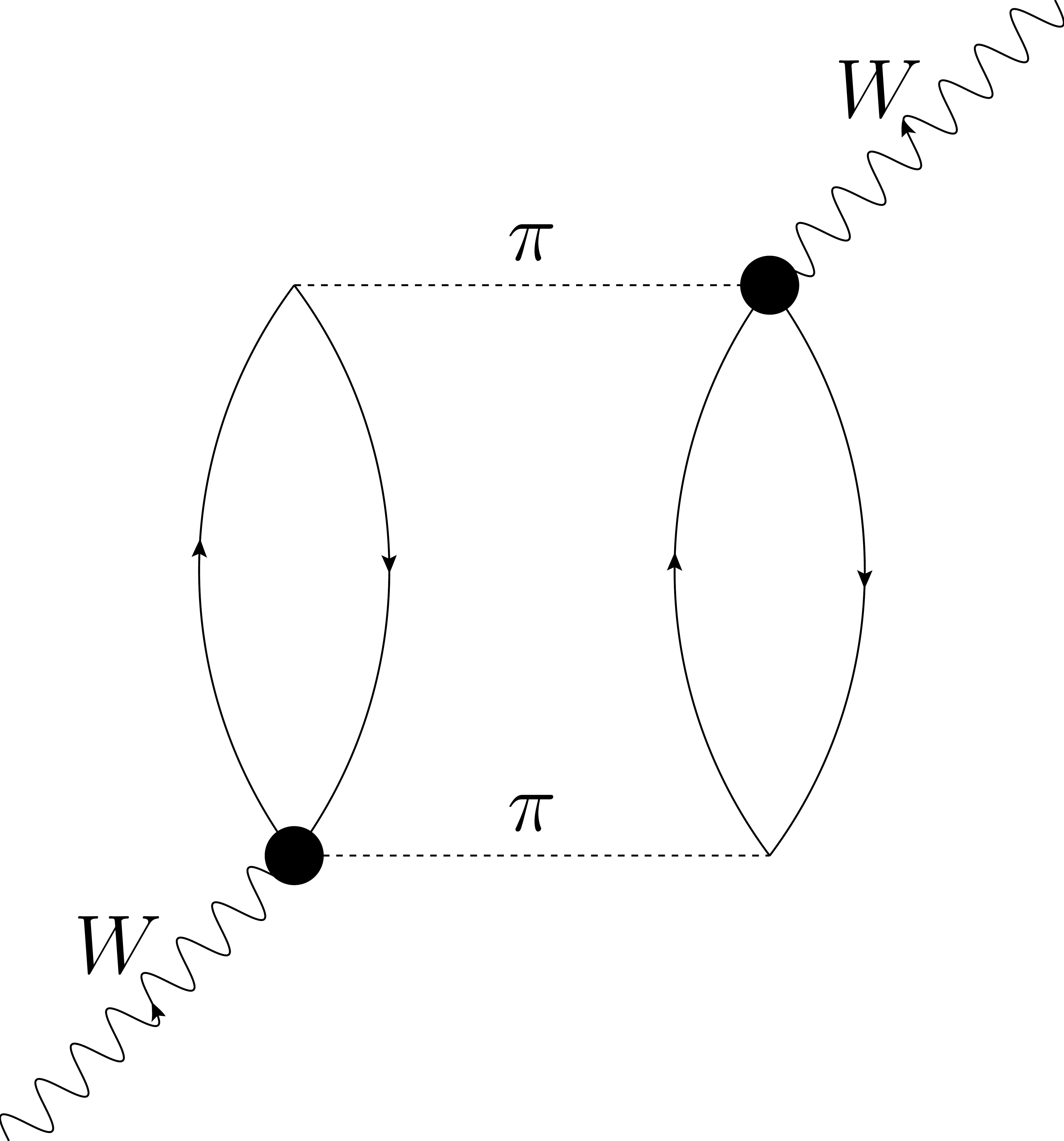}
\includegraphics[scale=0.1]{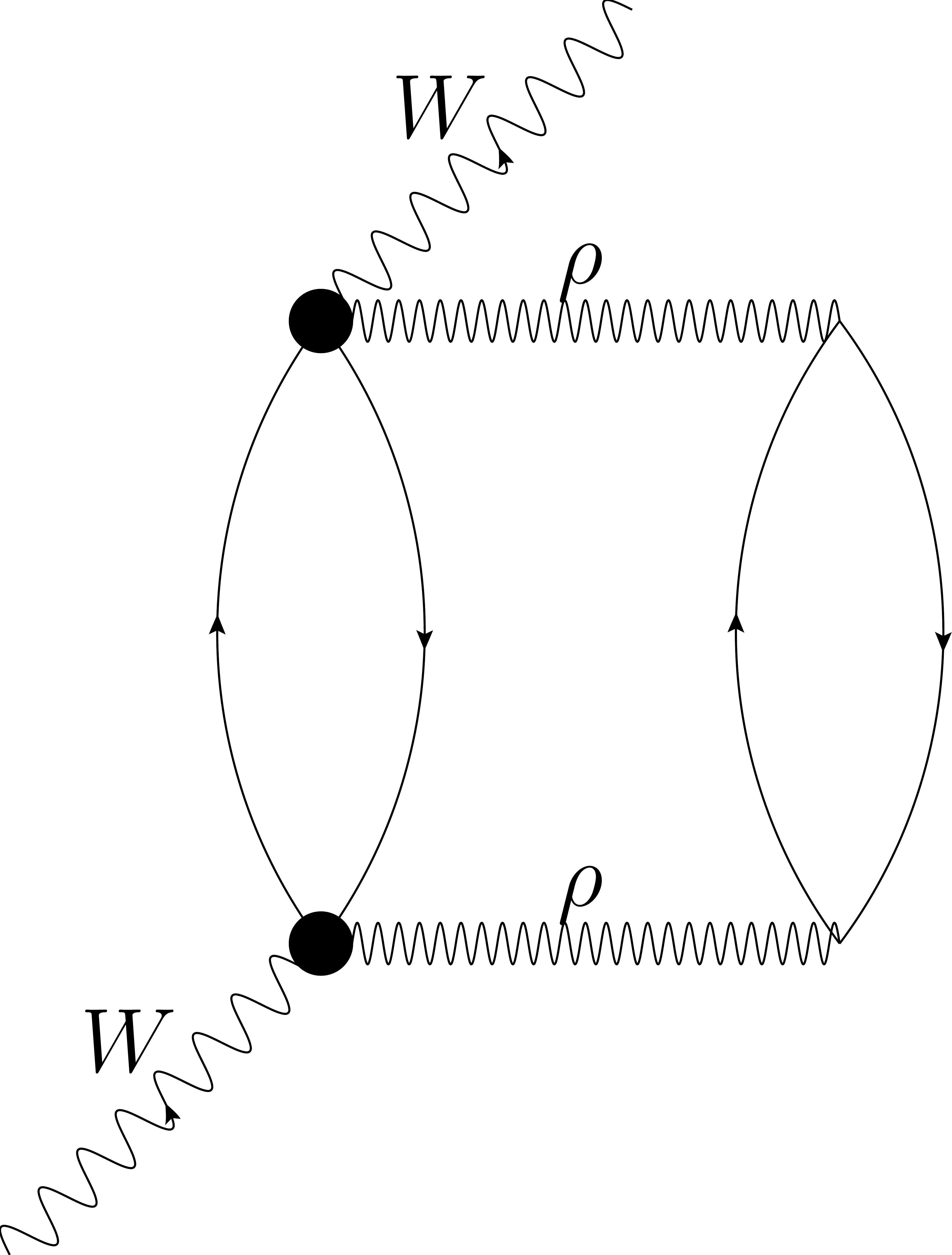}
\caption{Different 2p2h contributions to the $W^+$ self-energy in the nuclear matter driven by pion and $\rho$ exchanges, and which are taken into account in our work in a first step. The full circles account for the $WN\to \pi N$ and $WN\to \rho N$ amplitudes~\cite{Nieves:2011pp}.}
\label{fig2:w_se}
\end{figure}

The approach makes use of the local density approximation, in which  the nucleus is  locally treated  as the nuclear medium of constant density, to obtain results for finite nuclei from nuclear matter calculations. In one of the most important reaction mechanisms at intermediate energy-transfers, the mediator of electroweak interactions, the $W^\pm$ boson for CC interactions, traveling through the nuclear environment is absorbed by a pair of nucleons, producing another two.
We will call this mechanism 2p2h.\footnote{In the past, under this label we also referred to the absorption by three nucleons since it is included into the $\Delta$ self-energy. Here, we will treat this mechanism separately and denote it as 3p3h.} 
The interaction between two particle-hole excitations (ph-ph), in the spin-isospin channel,  can be separated into the longitudinal and the transverse channels. These are triggered by one pion and $\rho$ exchanges, respectively; with in-medium corrections which strongly influence both channels. To be precise, in a first step the many-body diagrams  calculated are shown in Fig.~\ref{fig2:w_se}. The structure of the $W^\pm N \to \pi N$ and $W^\pm N\to \rho N$ amplitudes can be found in Refs.~\cite{Nieves:2011pp,Hernandez:2016yfb}. It is worth mentioning that recently an extensive comparison of the electroweak pion production in the DCC (dynamical coupled channel) model~\cite{Kamano:2013iva,Matsuyama:2006rp}, Sato-Lee model~\cite{Sato:1996gk,Sato:2000jf,Sato:2003rq,Sato:2009de}, and the model of Refs.~\cite{Nieves:2011pp,Hernandez:2016yfb} has been performed~\cite{Sobczyk:2018ghy}. It has been shown that the latter approach -- in spite of its simplicity -- recovers the bulk of physical properties in the kinematical region of $\Delta(1232)$ excitation. This fact may serve as a confirmation that the model for the $W^\pm N \to \pi N$ reaction used in the present work for the 2p2h calculation is trustworthy.

We will focus on the inclusive nuclear reaction 
\begin{equation}
\nu_l (k) +\, A_Z \to l^- (k^\prime) + X 
\label{eq:reac}
\end{equation}
driven by the electroweak CC. The double differential cross section, with
respect to the outgoing lepton kinematical variables, for the process
of Eq.~(\ref{eq:reac}) is given in the Laboratory (LAB) frame by
\begin{equation}
\frac{d^2\sigma_{\nu l}}{d\Omega(\hat{k^\prime})dE^\prime_l} =
\frac{|\vec{k}^\prime|}{|\vec{k}~|}\frac{G^2}{4\pi^2} 
L_{\mu\sigma}W^{\mu\sigma} \label{eq:sec}
\end{equation}
with $\vec{k}$ and $\vec{k}^\prime~$ the LAB lepton momenta, $E^{\prime}_l =
(\vec{k}^{\prime\, 2} + m_l^2 )^{1/2}$ and $m_l$ the energy and the
mass of the outgoing lepton, $G=1.1664\times 10^{-11}$ MeV$^{-2}$, the
Fermi constant and $L$ and $W$ the leptonic and hadronic tensors,
respectively.  The leptonic tensor is given by\footnote{We
take $\epsilon_{0123}= +1$ and the metric $g^{\mu\nu}=(+,-,-,-)$.}:
\begin{eqnarray}
L_{\mu\sigma}&=& L^s_{\mu\sigma}+ {\rm i} L^a_{\mu\sigma} =
 k^\prime_\mu k_\sigma +k^\prime_\sigma k_\mu
- g_{\mu\sigma} k\cdot k^\prime + {\rm i}
\epsilon_{\mu\sigma\alpha\beta}k^{\prime\alpha}k^\beta \label{eq:lep}.
\end{eqnarray}
The hadronic tensor corresponds to the charged electroweak transitions of the
target nucleus, $i$, to all possible final states.  Expressions for the hadron tensor corresponding to the 2p2h diagrams in Fig.~\ref{fig2:w_se} can be found in Eqs. (27), (35), (36), (40) of Ref.~\cite{Nieves:2011pp}. However,  in the calculations presented in this work we will not perform an average over the initial nucleons momenta that appear in the electroweak amplitudes,  as it was done in Ref.~\cite{Nieves:2011pp} (see the discussion around  Eqs.~(18) and (19) in the latter reference). As was explained there, the difference between this approximated calculation and the full one is not large for inclusive cross sections. Here, however, we want to analyze exclusive hadronic final states which might be more sensitive to the averages done in the integrations. Let us also notice that in this way we automatically include the integration of Eq.~(31) of Ref.~\cite{Nieves:2011pp} which was introduced to deal with the pole of nucleon propagator (see the discussion above Eq.~(31) of Ref.~\cite{Nieves:2011pp}).

In the following, we will summarize several additional updates which we introduce to improve on the previous calculation.

%
%
%
%
\subsection{In-medium baryon-baryon effective interaction in the spin-isospin channel}
\label{sec:effective_int}
The key point of the approach is the assumption that the interaction between ph($\Delta$h)-ph($\Delta $h) nuclear excitations in the spin-isospin channel is originated from the $\pi$ and the $\rho$ exchanges, and modified in the nuclear medium \cite{Oset:1987re}. The one pion exchange potential between two nucleons has a longitudinal character and is given in momentum space by
\begin{equation}
    V_\pi(p) = \frac{f_{\pi NN}^2}{m_\pi^2}F_{\pi}^2(p^2)\, \vec{p}\,^2 D_\pi(p^2)\,  (\vec{\sigma}_1 \cdot \hat{p})\ (\vec{\sigma}_2 \cdot \hat{p})\ \vec{\tau}_1 \cdot \vec{\tau}_2
\end{equation}
with $\hat{p} =\vec{p}/|\vec{p}\,|$ the unitary three momentum transfer, and $\sigma_i$ and $\tau_i$ ($i=1,2$), Pauli matrices acting on the spin and isospin nucleon degrees of freedom, respectively. In addition,   
\begin{equation}
 D_\pi(p^2) = \frac{1}{p^2-m_\pi^2+i\epsilon}\, , \ \ \ \ 
 F_{\pi}(p^2)=\frac{\Lambda_\pi^2-m_\pi^2}{\Lambda_\pi^2-p^2}\, , \ \ \ \ \ \Lambda_\pi=1200\ \text{MeV}\, , \ \ \ \ m_\pi=139\ \text{MeV}
\end{equation}
We introduced $F_{\pi}(p^2)$ form factor to account for off-shell effects  on the $\pi NN$ vertex  and $f_{\pi NN}^2/4\pi = 0.08$.
The potential $V_\pi(p)$ can be split into scalar and tensor parts:
\begin{equation}
     (\vec{\sigma}_1 \cdot \hat{p}) \ (\vec{\sigma}_2 \cdot \hat{p}) = \frac13 \vec{\sigma}_1\cdot \vec{\sigma}_2 + \frac13 S_{12}(\hat{p})
     \label{eq:long}
\end{equation}
with the tensor operator $S_{12}(\hat{p}) = 3 (\vec{\sigma}_1 \cdot \hat{p})\ (\vec{\sigma}_2 \cdot \hat{p})- \vec{\sigma}_1\cdot \vec{\sigma}_2$. 
The Fourier transform to the coordinate space, in the static limit ($p^0=0$) and neglecting the $F_\pi(p^2)$ form factor, of the scalar potential gives rise to
\begin{equation}
    V_\pi(\vec{r}\,) = \frac13 \frac{f_{\pi NN}^2}{m_\pi^2} \bigg(\delta^3(\vec{r})- \frac{m_\pi^2}{4\pi}\frac{e^{-m_\pi |\vec{r}|}}{|\vec{r}|} \bigg) \vec{\sigma}_1 \cdot \vec{\sigma}_2\ \vec{\tau}_1 \cdot \vec{\tau}_2
\end{equation}
with $|\vec{r}|\,$, the $NN$ relative distance. The term proportional to $\delta^3(\vec{r})$ comes from the construction of the potential when nucleons are treated as point-like particles. This is not a correct physical behaviour and it is firstly corrected by the form-factor $F_{\pi}(p^2)$. Nevertheless, it is  well known that the  strong short-range correlations prevent nucleons from getting close to each other, and thus  at shorter distances, two pion exchange mechanism gains on importance.  One pion exchange describes the long-range part of interaction, corresponding to distances $|\vec{r}\,|\ge \lambda_\pi = \frac{1}{m_\pi} \approx 1.4\ \text{fm}$.
Moreover, short-range correlations are modified inside of the nuclear medium.
 
\begin{figure}[h]
\centering
\includegraphics[scale=0.8]{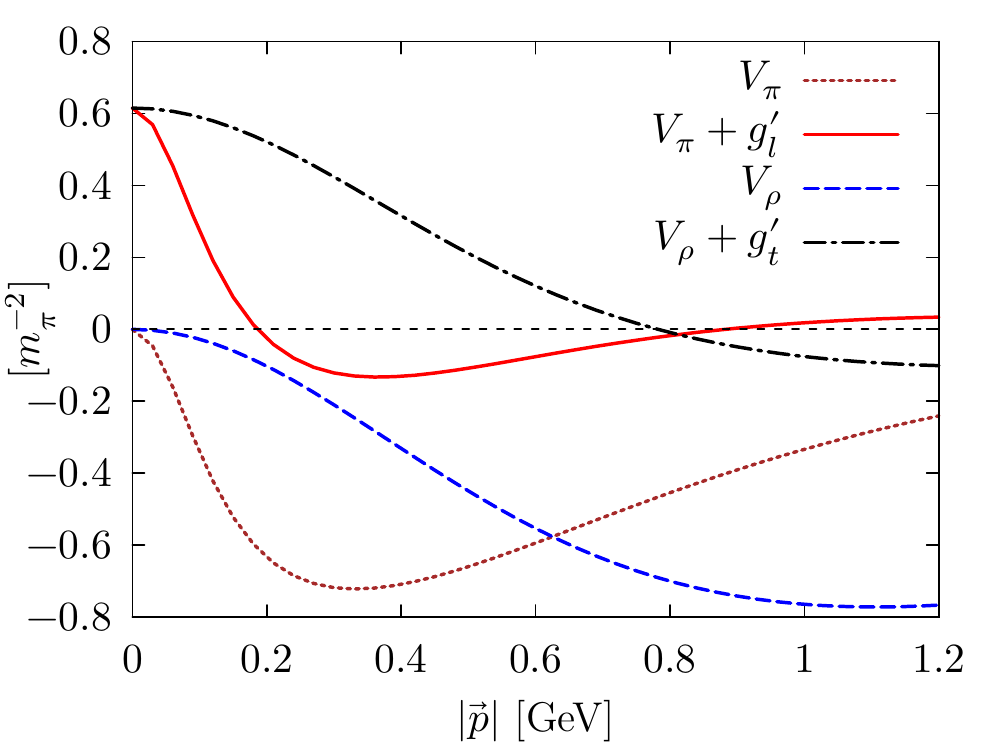}
\includegraphics[scale=0.8]{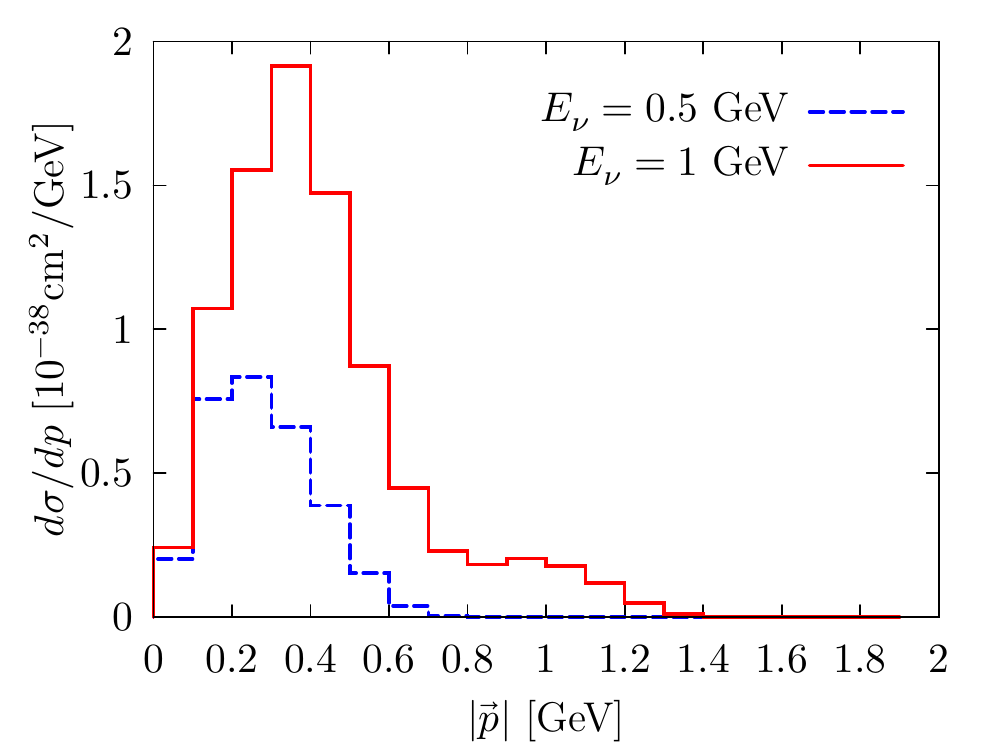}

\caption{Left panel: longitudinal and transverse channels of in-medium $\sigma\sigma\tau\tau$ potential $V(p)$ in the static limit $p^0=0$. Solid red  and dotted-dashed black lines correspond to the final interactions, with the addition of the $g_l'$ and $g_t'$ terms.  Right panel: Monte Carlo distribution of momentum exchanged between two ph-ph excitations in the calculation of the 2p2h reaction mechanism, for incoming neutrino energies of 0.5 and 1 GeV.}
\label{fig:potential}
\end{figure}

 In addition, the vector-isovector channel of the $NN$ interaction is also strongly influenced by the  $\rho-$meson (transverse) exchange:
\begin{equation}
    V_\rho(p) = C_\rho\frac{f_{\pi NN}^2}{m_\pi^2}F_{\rho}^2(p^2)\, \vec{p}\,^2 D_\rho(p^2)\, (\vec{\sigma}_1 \times \hat{p}) \ (\vec{\sigma}_2 \times \hat{p})\ \vec{\tau}_1 \cdot \vec{\tau}_2
\end{equation}
with 
\begin{equation}
 D_\rho(p^2) = \frac{1}{p^2-m_\rho^2+i\epsilon}\, , \ \ \ \ 
 F_{\rho}(p^2)=\frac{\Lambda_\rho^2-m_\rho^2}{\Lambda_\rho^2-p^2}\, , \ \ \ \ \ \Lambda_\rho=2500\ \text{MeV}\, , \ \ \ \ m_\rho=770\ \text{MeV}\, , \ \ \ \ \ C_\rho=2
\end{equation}
Again, it can be separated into the scalar and tensor parts,
\begin{equation}
     (\vec{\sigma}_1 \times \hat{p}) \ (\vec{\sigma}_2 \times \hat{p}) = \frac23 \vec{\sigma}_1\cdot \vec{\sigma}_2 - \frac13 S_{12}(\hat{p})
     \label{eq:trans}
\end{equation}
with the former one also giving rise to a non-regularized $\delta^3(\vec{r}\,)$ term. On the other hand,  the tensor components have an opposite sign for longitudinal and transverse parts and thus partially cancel (compare Eqs.~\eqref{eq:long} and \eqref{eq:trans}). The approach of describing the nucleon-nucleon interaction in terms of meson-exchanges breaks down at short distances where the potential is known to be strongly repulsive.\footnote{In the calculations presented in this work the main strength will come from low and medium values of transferred momenta $\vec{p}$ between nucleons (for neutrinos $E_\nu\le 2$ GeV, $|\vec{p}\,|$ peaks below $0.5$ GeV, see the right panel in Fig.~\ref{fig:potential}). These values of momenta probe mainly the region described by one and two pion exchange potential.} 
At this stage we follow Ref.~\cite{Oset:1987re} and introduce the effective  terms $g_l'$ and $g_t'$ to account for  short-range effects of the $NN$ $\sigma\sigma\tau\tau$ interaction inside of the nuclear medium,  
\begin{equation}
    V(p) = V_\pi(p) + V_\rho(p) + \frac{f_{\pi NN}^2}{m_\pi^2}\, g_l'(p)\, (\vec{\sigma}_1 \cdot \hat{p})\ (\vec{\sigma}_2 \cdot \hat{p})\ \vec{\tau}_1 \cdot \vec{\tau}_2 + C_\rho\frac{f_{\pi NN}^2}{m_\pi^2}\, g_t'(p)\, (\vec{\sigma}_1 \times \hat{p})\ (\vec{\sigma}_2 \times \hat{p})\ \vec{\tau}_1 \cdot \vec{\tau}_2
\end{equation}
To obtain $g_l'$ and $g_t'$, we follow the discussion of Ref~\cite{Oset:1979bi}, where a phenomenological correlation function $\Omega(\vec{r}\,)$ was introduced, fulfilling the conditions:
\begin{equation}
    \Omega(\vec{r}=\vec{0}\,) = 0\, , \ \ \ \ \ \Omega(\vec{r}\,)\approxeq 1\ \text{for}\ |\vec{r}\,| \gtrapprox r_c\, .
\end{equation}
with $r_c$  a distance between nucleons below which they feel a strong repulsion whose  details  cannot be disentangle in the medium. The distance $r_c$ is estimated to be around $2.6$ fm, which corresponds to the mass of the $\omega$ meson, $q_c=780$ MeV. 
The desired behaviour of the potential is imposed by modulating the potential with the short-distance function,
\begin{equation}
V(\vec{r}\,) = \big[    V_\pi(\vec{r}\,) + V_\rho(\vec{r}\,) \big]\, \Omega(\vec{r}\,)
\end{equation}
The functional form of $\Omega(\vec{r}\,)$ is taken to be $[1-j_0(q_c|\vec{r}\,|)]$ (spherical Bessel function) whose Fourier transform reads:
\begin{equation}
    \Omega(\vec{k}) = (2\pi)^3 \delta^3(\vec{k}) - \frac{2\pi^2}{q_c^2} \delta(|\vec{k}|-q_c) \, .
\end{equation}
With all these elements we can write the potential in momentum space as:
\begin{align}
   V(p)& = \int \frac{d^3k}{(2\pi)^3} \left(V_\pi(p_0, \vec{p}-\vec{k}) + V_\rho(p_0, \vec{p}-\vec{k}) \right) \Omega(\vec{k})\nonumber \\
   &=V_\pi(p) + V_\rho(p) - \int \frac{d^2 \hat{k}}{4\pi} \left(V_\pi(p_0,\vec{p}-q_c\hat{k}) + V_\rho(p_0,\vec{p}-q_c\hat{k}) \right) 
\end{align}
After performing the integration, averaging over angles, we finally obtain~\cite{Oset:1987re}:
\begin{align}
&g_l'(p) = -\left(\vec{p}\,^2+\frac13\, q_c^2\right)\, \tilde{F}_\pi^2 \,\tilde{D}_\pi - \frac23 C_\rho \,q_c^2\,\tilde{F}_\rho^2\, \tilde{D}_\rho \, ,\nonumber \\
 &g_t'(p) = -\frac13\, q_c^2\, \tilde{F}_\pi^2\, \tilde{D}_\pi -\left(\vec{p}\,^2+ \frac23 q_c^2\right)\, C_\rho\,\tilde{F}_\rho^2\, \tilde{D}_\rho\, .
\end{align}
where $\tilde{F}_{\pi,\rho}$ and $\tilde{D}_{\pi,\rho}$ correspond to form-factors and propagators with shifted momentum $\vec{p}\,^2 \to \vec{p}\,^2+q_c^2$.
In the left panel of Fig.~\ref{fig:potential} we show the effect of $g_l'$ and $g_t'$ on the potential.
Let us notice that in the limit of high momentum transfer $|\vec{p}\,|\gg q_c$, the potential goes to $0$, while in the regime of low momentum transfers we obtain $g_l'=g_t'=0.63$. This is a value we used in previous calculations of the RPA~\cite{Nieves:2004wx, Nieves:2017lij} and 2p2h excitation~\cite{Nieves:2011pp} effects on CCQE reactions.
In the right panel of Fig.~\ref{fig:potential} we show the region of $|\vec{p}\,|$ (exchanged momentum between two ph-ph excitations) probed in the calculation of 2p2h contribution. For both neutrino energies, $E_\nu=0.5$ GeV and  $E_\nu=1$ GeV, the momentum distributions peak at rather low values (0.2 and 0.4 GeV respectively). In this region the contribution coming from the longitudinal part of interaction $V_l(p) = V_\pi(p) + g_l'(p)$ is relatively small in comparison to the transverse one (see the left panel of Fig.~\ref{fig:potential}). The short-distance $g_l'$ term largely interferes with the pion-exchange potential giving rise to an interaction significantly different than $V_\pi$. The former term modifies the low energy region, but also cancels out the high-energy contribution that would otherwise emerge from one pion exchange. On the other hand, the $\rho$ propagator is largely suppressed by $m_\rho$, but the transverse interaction gets enhanced due to $g_t'$. Ultimately, the strength is dominated by the transverse channel, shown in Fig.~\ref{fig:potential} by the dotted-dashed black line.

An analogous discussion to that above holds in the case of $\Delta$h-$\Delta$h and  ph-$\Delta$h effective interaction in the nuclear medium, with appropriate spin and isospin operators $\vec{\sigma} \to \vec{S}$ and $\vec{\tau} \to \vec{T}$ and  replacing the coupling constant $f_{\pi NN}$ by $f_{\pi N\Delta}^*$ ($=\sqrt{4\pi \times 0.36}$). In addition, $C_\rho$ and the form-factors $F_\pi$, $F_\rho$ are the same for both nucleon and $\Delta(1232)$ cases~\cite{Oset:1987re,Oset:1979bi}.

%
%
%
%
\subsection{Treatment of $\Delta(1232)$}
\label{sec:delta_treatment}
\begin{figure}[h]
\centering
\includegraphics[scale=0.1]{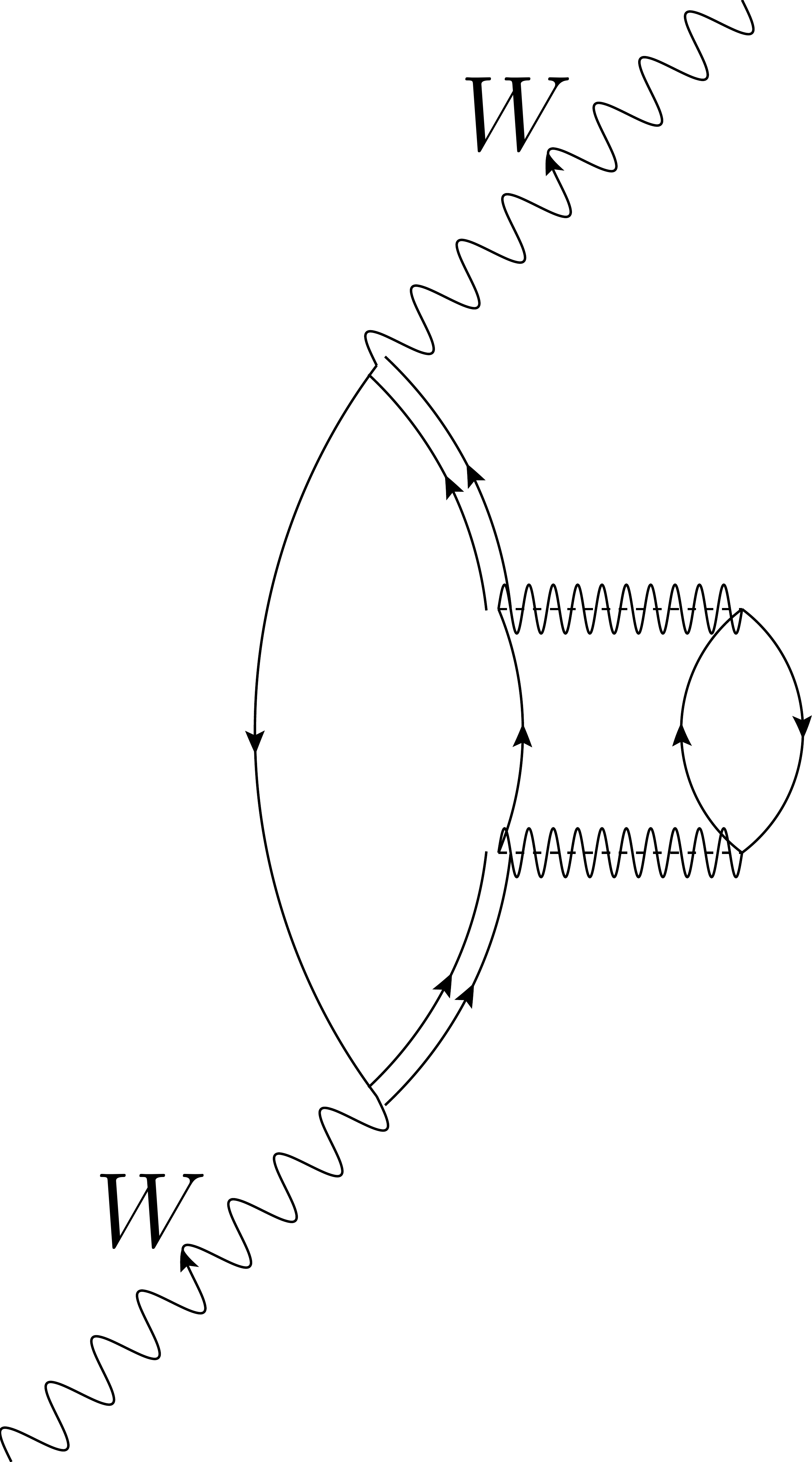}
\hspace{2 cm}
\includegraphics[scale=0.1]{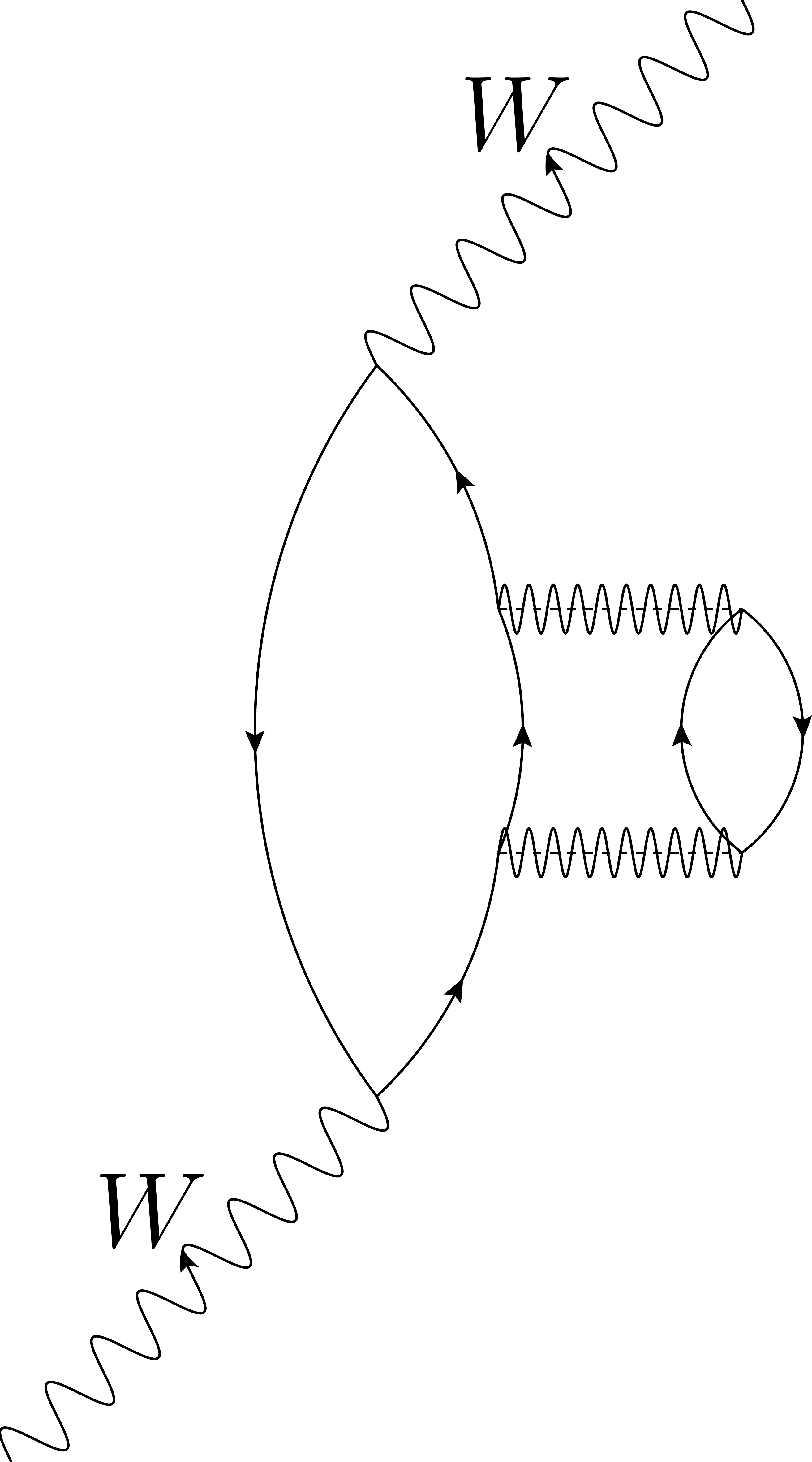}
\hspace{2 cm}
\includegraphics[scale=0.1]{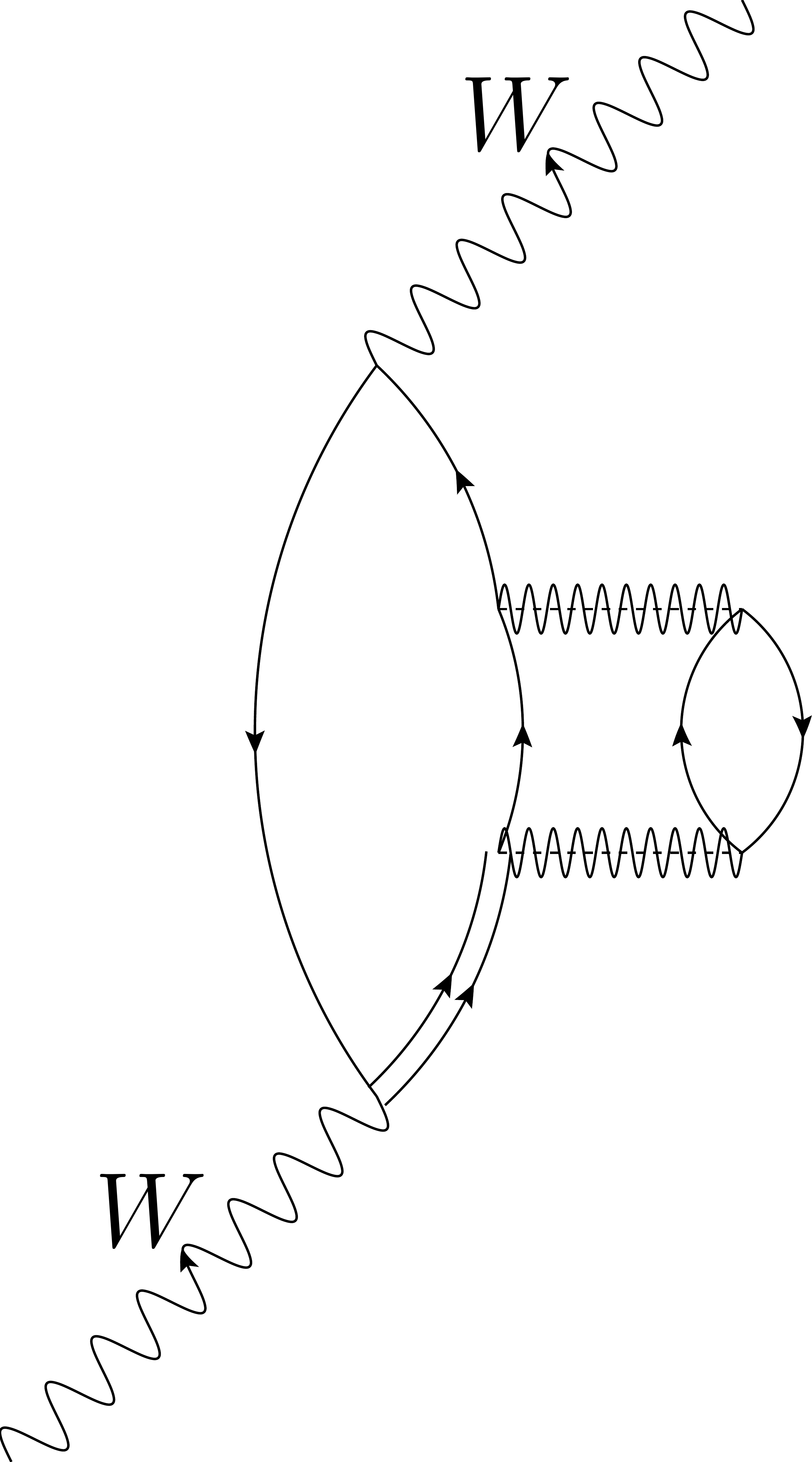}
\caption{Three different contributions to the 2p2h hadron tensor. In the left plot the diagram has two direct $W^+ \Delta N$ vertices (and two $\Delta$ propagators) and it can be cast as $\Delta $h excitation, with a self-energy insertion for the resonance. We will refer to it as ``$\Delta\Delta$-diagram''. In the middle, the diagram contains two direct $W^+NN$ vertices, and it builds an in medium  nucleon-selfenergy.  In the right plot an interference term is shown with both $W^+\Delta N$ and $W^+ NN$ vertices (one $\Delta$ and another nucleon propagators), referred to as ``$N\Delta$-diagram''. In all the cases the second ph is excited  either from the  longitudinal or transverse spin-isospin interaction.}
\label{fig2:2_delta}
\end{figure}

We briefly sketch here the steps given in Ref.~\cite{Nieves:2011pp}   to calculate the inclusive 2p2h cross section:
\begin{enumerate}
    \item Performing the full calculation of diagrams implicit in Fig.~\ref{fig2:w_se}, replacing the pion and $\rho$ exchanges with the  effective spin-isospin interaction both in the longitudinal and transverse channels. Next, the contributions which have two direct $W^+N\Delta$ vertices (we will denote them as $\Delta\Delta$-diagrams, see the left panel of Fig.~\ref{fig2:2_delta}) or two $W^+NN$ vertices (the central panel of Fig.~\ref{fig2:2_delta}), for both longitudinal and transverse channels are subtracted from the full sum. The latter diagrams, driven by the longitudinal and transverse spin-isospin interactions, are removed because they give contribution to the nucleon spectral function, and thus should be taken into account when the QE mechanism is considered. In turn, the  $\Delta\Delta$-diagrams can be seen as $\Delta$h nuclear excitation, with resonance dressed in the nuclear medium. 
    
    \item Computing the $\Delta $h excitation by the $W-$boson, including the imaginary part of the $\Delta-$self energy,  which  accounts for  the 2p2h and 3p3h contributions to the $\Delta$ self-energy, $\Sigma_\Delta$, as calculated in Ref.~\cite{Oset:1987re}. There are also some pion production quasi-elastic contributions included to $\text{Im}\Sigma_\Delta$, which however do not lead to multi-nucleon absorption in first approximation. Note that in addition in Ref.~\cite{Nieves:2011pp}, $\text{Re}\Sigma_\Delta$ is set  to zero since its  inclusion would require a detailed RPA re-summation, with separate longitudinal and transverse series.     
\end{enumerate}

To find the predictions of the model of Ref.~\cite{Nieves:2011pp} for exclusive observables (such as outgoing nucleons momenta), we encounter a fundamental obstacle in the aforementioned procedure. The $\text{Im}\Sigma_\Delta$, which gives a major contribution to the total 2p2h cross section, is parametrized in Ref.~\cite{Oset:1987re} in terms of the kinetic energy of a pion that would excite a $\Delta$ with the corresponding invariant mass. The information about underlying dynamics of excited nucleons is already integrated out. To overcome this problem, we have explicitly evaluated here the $\Delta\Delta$-contribution (the first diagram of Fig.~\ref{fig2:2_delta}), instead of using the pre-computed Im$\Sigma_\Delta$ with the appropriate $C_{A_2}$ parameter (see Eq. (51) of Ref.~\cite{Nieves:2011pp}). We have followed the original calculation of the $\Delta$ self-energy carried out in Ref.~\cite{Oset:1987re} where $g'_l(p)\neq g'_t(p) \neq g'$, just as explained in the previous subsection. The energy dependence of $g'_l$ and $g'_t$ parameters will affect especially high-energy transfers regime where particularly $g_l'$ differs considerably from the constant $g'=0.63$.

%
%
%
%
\subsection{Further refinements of $\Delta$}
\label{sec:refinements}
Following the results presented in Ref.~\cite{Hernandez:2016yfb}, we will introduce two modifications to the treatment of $\Delta$ excitation which were proposed to achieve a better agreement with the neutrino ANL and BNL data in $\nu_\mu n\to \mu^- n \pi^+$ channel. First of all, we will employ $C_A^5$ form-factor for the $\Delta - N$ electroweak transition given by
\begin{equation}
    C_{A}^5(q^2) = \frac{1.18}{(1-q^2/M_{A\Delta}^2)^2}\, , \ \ \ \ \ M_{A\Delta} = 950\ \text{MeV}\,.
\end{equation}
Secondly, we  change the $\Delta$ propagator from the Rarita-Schwinger form to the pure spin-$3/2$ projector operator, including a $p^2_\Delta/M^2_\Delta$ local factor which is equal to one at the $\Delta-$peak,
\begin{align}
&    G_{\mu\nu}(p_\Delta) = \frac{p^2_\Delta}{M^2_\Delta} \frac{P_{\mu\nu}^{\frac32}(p_\Delta)}{p_\Delta^2-M_\Delta^2+i M_\Delta \Gamma_\Delta} \nonumber \\
&    P_{\mu\nu}^{\frac32}(p_\Delta) = -(\slashed{p}_\Delta+M_\Delta) \bigg[ g_{\mu\nu} - \frac{1}{3} \gamma_\mu \gamma_\nu - \frac{1}{3p_\Delta^2} (\slashed{p}_\Delta \gamma_\mu p_{\Delta\nu}+p_{\Delta\mu} \gamma_\nu \slashed{p}_\Delta)   \bigg]
\end{align}
This corresponds to the prescription of using consistent $\Delta$ couplings~\cite{Pascalutsa:2000kd}, however it does not precisely recover the results of Ref.~\cite{Hernandez:2016yfb},  where a $1/2$-spin part of $\Delta$ propagator was partially retained. We have checked, though, that the difference is small and in our calculation can be safely neglected.

Furthermore, the neglect of the spin$-1/2$ degrees of freedom  and the use of consistent vertices lead to an accurate reproduction of the Watson's unitary theorem in the dominant $P_{33}-$multipole~\cite{Alvarez-Ruso:2015eva}, without  including any further sizable phase~\cite{Hernandez:2016yfb}.

%
%
%
%
\subsection{3p3h}
\label{sec:3p3h}
The new treatment of the $\Delta\Delta$-diagram, as exposed in Subsec.~\ref{sec:delta_treatment}, enables us to calculate the distribution of outgoing nucleons for the two body reaction mechanism associated to this diagram. In the former works \cite{Nieves:2011pp, Nieves:2012yz}, the imaginary part of the $\Delta$ self-energy was comprised of four contributions (Pauli blocking, 1p1h1$\pi$, 2p2h and 3p3h) out of which the latter two were included to account for the multi-nucleon knockout  inclusive cross section. Presently, the MC generators which use this model usually follow this prescription.
This approximation is reasonable since we suspect that for three outgoing nucleons the least energetic one will hardly ever be detected due to the experimental energetic threshold.

 In the following, we will treat separately the 2p2h and 3p3h contributions, since we are interested in the dynamics of the outgoing nucleons. The only 3p3h source in the model comes from the corresponding mechanism included in  $\text{Im}\Sigma_\Delta$ in the $\Delta\Delta$-diagram. However, we have many other sources of two body absorption besides the latter diagram, some of them not involving the excitation of the $\Delta$ resonance. 
For the time being, we use a parametrization of the 3p3h process reported in Ref.~\cite{Oset:1987re} where it is given as a function of the kinetic energy of a pion that would excite a $\Delta$ with the corresponding invariant mass. 
The calculation of the final states in the 3p3h process is a non-trivial task and goes beyond the scope of the present analysis.

%
%
%
%
\section{Calculation in NEUT}
\label{sec:NEUT}

 In NEUT~\cite{Hayato:2009zz}, the events are generated according to the distribution of the outgoing lepton, i.e. using the weight given by the value of double-differential inclusive cross section, expressed as the contraction of lepton  and hadron tensors as given by  Eq.~\eqref{eq:sec}. The hadron tensor $W^{\mu\nu}$ is evaluated following Ref.~\cite{Nieves:2011pp}. It is computed separately for proton-neutron and proton-proton final states and used to provide isospin dependent final states.  The location of the interaction vertex in the nucleus is chosen according to the density profile and the initial state nucleons are picked below the Fermi level corresponding to the radial position following the local Fermi gas model recipe.  
 The outgoing nucleons at the weak vertex are distributed according to the  available phase-space. They are generated uniformly in the center of the mass of the hadronic system and boosted to the laboratory rest frame. Next, their momenta are tested against the local Fermi level to implement Pauli blocking. This procedure neglects the dynamics of underlying nuclear model and produces a symmetric distribution of outgoing nucleons.  The produced pair of nucleons is fed into the NEUT cascade model accounting for the transport of nucleons in the high density nuclear medium. The model of the NEUT cascade is based on the model developed in \cite{Salcedo:1987md} modified by the experimental nucleon-nucleon and pion-nucleon cross section data when available \cite{PinzonGuerra:2018rju}.
The same cascade model is used to calculate the final state nuclear re-interaction migration matrices that are used to estimate the effects on the new 2p2h hadron kinematics predictions. These migration matrices describe the transport of the particles from the interaction point to the outside of the nucleus one by one and considering the momentum and nature of the particles. These matrices are used later to compute the visible final energy outside of the nucleus, see the results of Sec.~\ref{sec:available}.

%
%
%
%
\section{Results}
\label{sec:results}
As we have stated in Subsec.~\ref{sec:3p3h}, in the following analysis we will treat separately the 3p3h contribution to the $\Delta$ self-energy, while the NEUT results are shown with both 2p2h and 3p3h inclusions according to the current model implementation.

\subsection{Total cross section and lepton differential distributions}
\begin{figure}[h]
\centering
\includegraphics[scale=0.8]{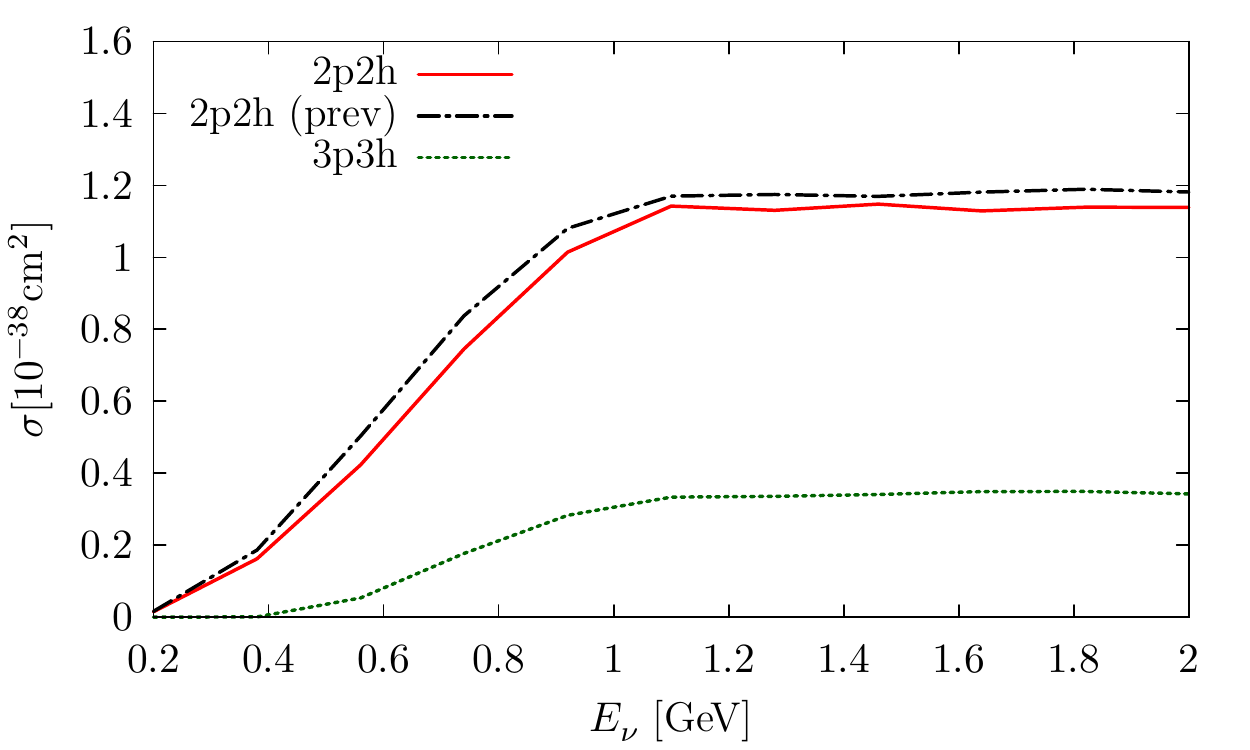}
\caption{ Total cross section $\sigma$ $[10^{-38} {\rm cm}^2]$ on $^{12}$C as a function of incoming neutrino energy $E_{\nu}$. The solid red and dashed-dotted black curves correspond to two calculations of the 2p2h explained in the text. The difference between both cross sections can be understood as a theoretical uncertainty of our approach. The dotted green curve corresponds to 3p3h and should be added to the 2p2h contribution. In all cases, the  cut  $\mid\vec{q}\mid < 1.2$ GeV in the momentum transfer is applied. 
}

\label{fig:total}
\end{figure}

In Fig.~\ref{fig:total} we show the total cross section for both 2p2h and 3p3h mechanisms on $^{12}$C. With the solid red line we plot the result obtained with the present calculation. The dashed-dotted black line labeled ``2p2h (prev)'' corresponds to the 2p2h result in which we include ${\rm Im}\Sigma_\Delta$ as in Ref.~\cite{Nieves:2011pp}.\footnote{It means that we do not follow modifications described in Subsec.~\ref{sec:delta_treatment} to compute the $\Delta-\Delta$ contribution, however, we do introduce the refinements described in Subsecs.~\ref{sec:effective_int}, \ref{sec:refinements}.} We interpret the difference between the two calculations as a theoretical uncertainty of our approach, which is at the order of $5-10\%$. In the latter one, the 2p2h inclusion into the $\Delta$ self-energy was parametrized (thus introducing some approximations).
The 3p3h contribution, calculated according to Ref.~\cite{Oset:1987re}, is shown by the green dotted line. It corresponds to around $20\%$ of the total 2p2h cross section for neutrino energies above 1 GeV. This prediction should be treated with some caution. In principle a more careful 3p3h calculation could be performed but, as mentioned, it is technically complicated and the smallness of the expected differences  discourage for the time being from further inquiries. Still, the approach we follow is one of the very few existing and the most widely used in the studies of neutrino induced reactions.

All the curves in Fig.~\ref{fig:total} have a similar energy dependence, with a plateau above $E_\nu\approx 1.1$ GeV. The reason for this behaviour is a cut we impose in the momentum transfer $|\vec{q}\,|<1.2$ GeV, which affects largely the cross section for $E_\nu>1$ GeV. Thus, for instance, for $E_\nu=1.5$ GeV this cut is responsible for neglecting around $10\%$ of the available phase space. 
Our model cannot probe high energy momentum transfers for various reasons. First of all, the non-resonant terms of $W^\pm\pi N$ vertex are obtained within the chiral perturbation theory and can be safely used only for low and moderate energy-momentum transfers. Besides, we do not include higher resonances above $\Delta$ exchange. Moreover, our model for effective in-medium interactions was constructed to describe exchange of moderate momenta between ph or $\Delta$h excitations. This momentum transfer sharp-cutoff corrects for the growth of the 2p2h+3p3h cross section with the neutrino energy, for $E_\nu > 1.2$ GeV,  found in the previous results of \cite{Nieves:2011pp}  (see for instance  green squares in Fig.5 of Ref.~\cite{Nieves:2012yz}). Below, $E_\nu< 1.2$ GeV, we successfully re-obtain the bulk of the results already published. There exist minor differences that can be ascribed to  the new treatment of the $\Delta$. We will see below that this is also the case for the outgoing lepton differential distributions. 

\begin{figure}[h]
\centering
\includegraphics[scale=0.46]{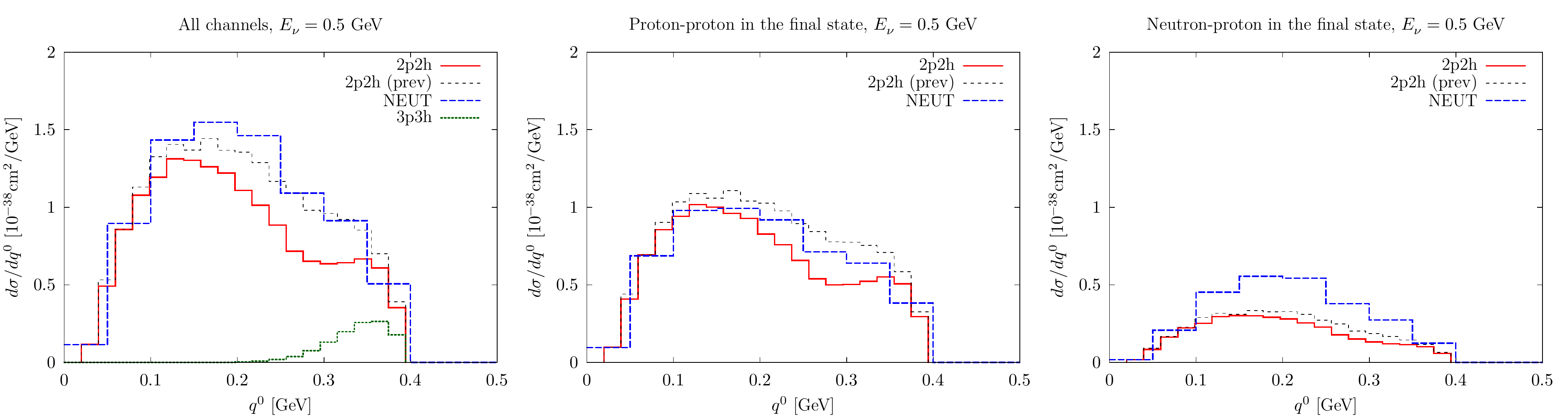}
\includegraphics[scale=0.46]{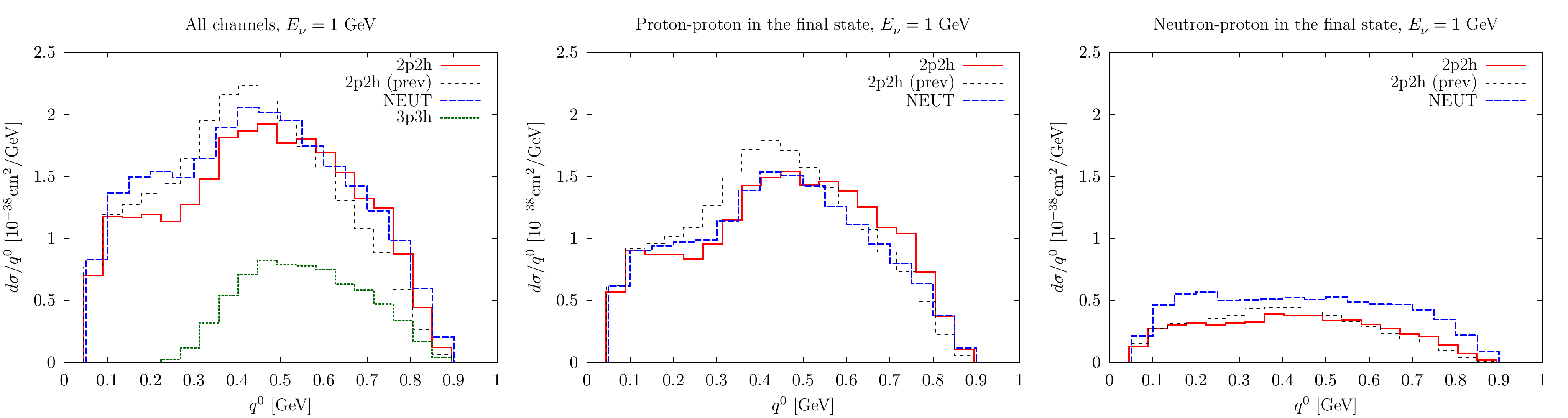}
\includegraphics[scale=0.46]{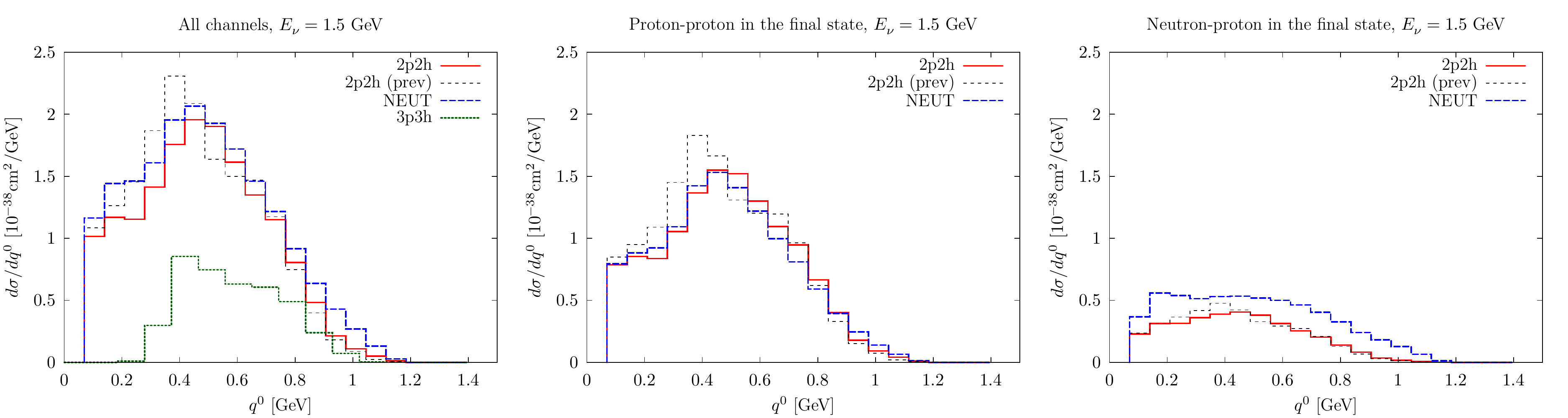}
\caption{$d\sigma/dq^0$ differential cross section on $^{12}$C for three different incoming neutrino energies: $E_\nu=0.5$, 1 and 1.5 GeV, displayed in the top, middle and bottom rows, respectively. Results in the left column are summed over isospin, while the central (right) column corresponds to two protons (one  neutron and one proton) in the final state. In all cases, the  cut  $\mid\vec{q}\mid < 1.2$ GeV in the momentum transfer is applied.  }
\label{fig:dsigmadE}
\end{figure}

\begin{figure}[h]
\centering
\includegraphics[scale=0.46]{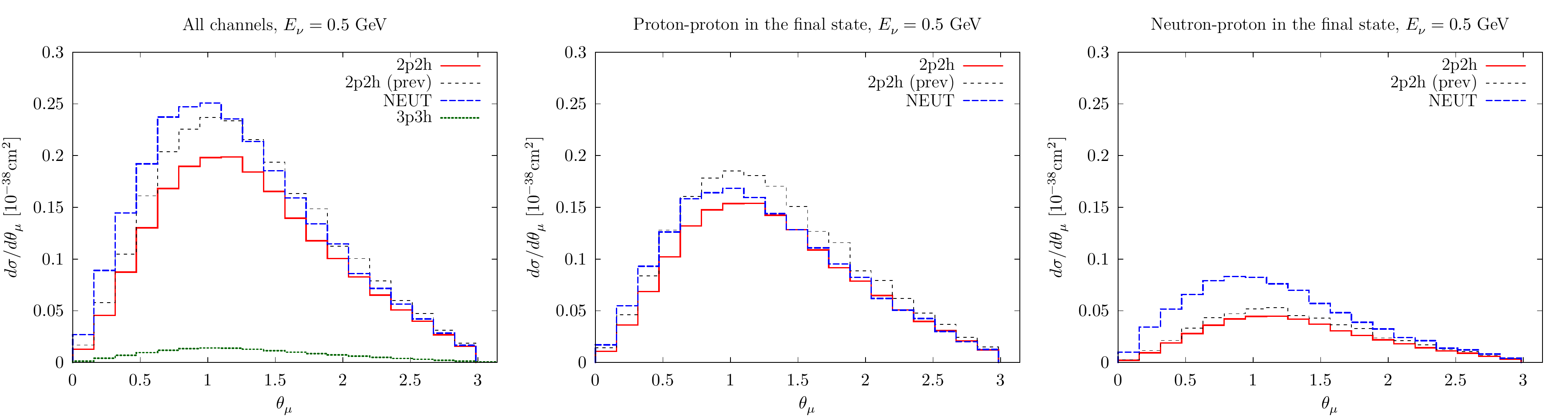}
\includegraphics[scale=0.46]{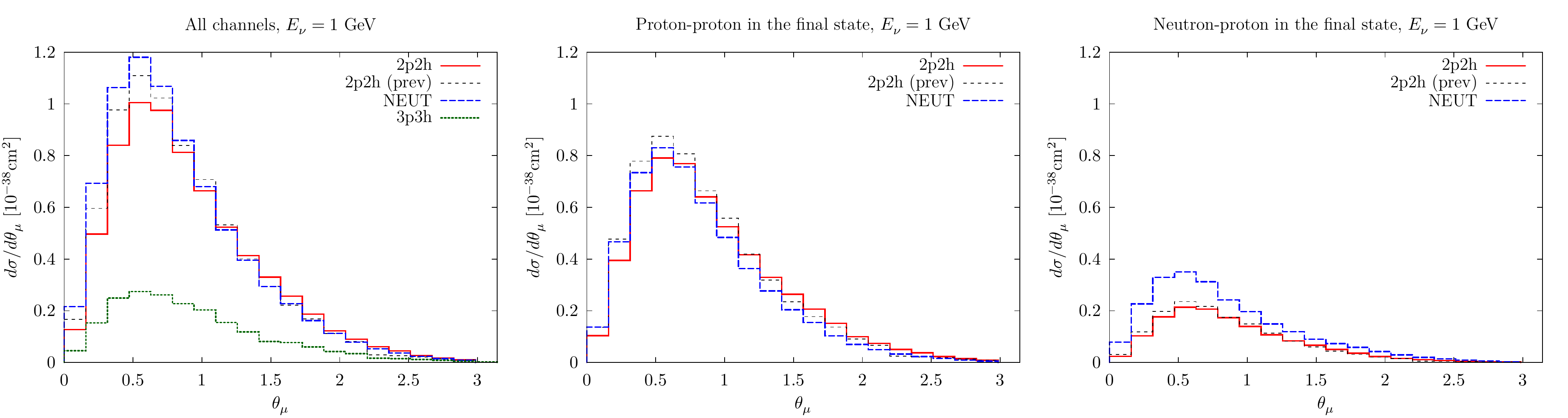}
\includegraphics[scale=0.46]{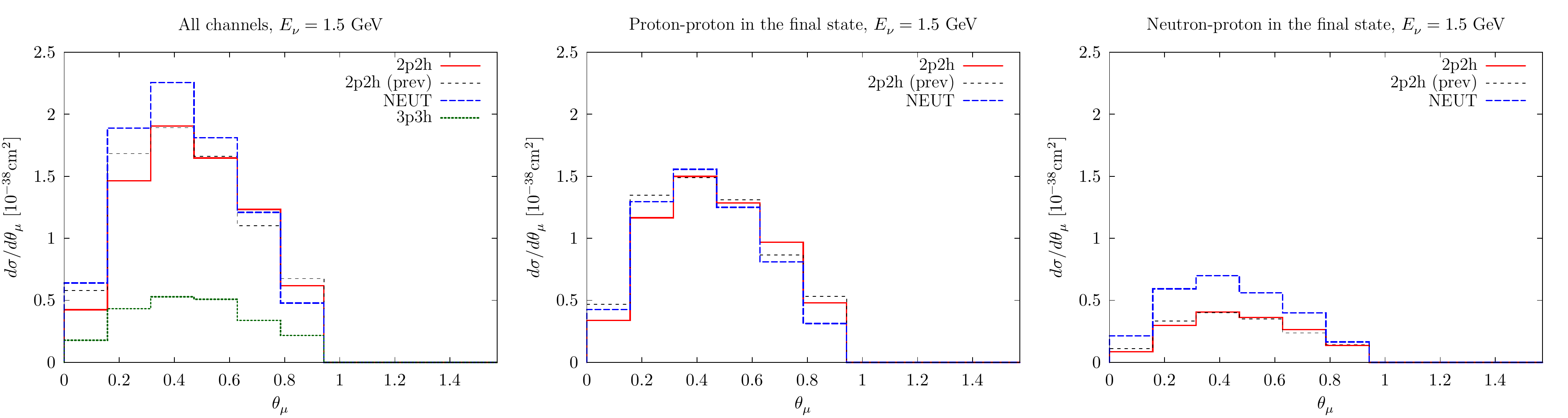}
\caption{As in Fig.~\ref{fig:dsigmadE},but for $d\sigma/d\theta_\mu$ differential cross section.
}
\label{fig:dsigmadTh}
\end{figure}

In Figs.~\ref{fig:dsigmadE} and \ref{fig:dsigmadTh} we show the $d\sigma/d q^0$ and $d\sigma/d\theta_\mu$ lepton differential cross sections respectively, for three incoming neutrino energies, $E_\nu=0.5,\,1,\, 1.5$ GeV (rows) and for three cases of the isospin state of final nucleons: either summed over isospin states (left), or  for two protons (middle), or for neutron-proton final state (right). 
The dashed blue line corresponds to the NEUT result, which follows the approach of Ref.~\cite{Nieves:2011pp} (we remind that in this case the 3p3h contribution is included).
For the $d\sigma/d q^0$ distribution at $E_\nu=1$ GeV (the middle row in Fig.~\ref{fig:dsigmadE}) we can clearly observe two peaks, at $q^0\simeq 0.15$ GeV and $q^0\simeq0.45$ GeV both for the NEUT and for the 2p2h distributions.  They correspond to two distinct dynamical mechanisms which dominate the total cross section. The first one (which peaks at lower energy transfers) is the contribution driven by the interference $N\Delta$-diagram (see the last plot in Fig.~\ref{fig2:2_delta}). The second one comes from the $\Delta\Delta$-diagram (first one in Fig.~\ref{fig2:2_delta}). 
Their relative position can be understood since a $\Delta$ resonance has a mass around $0.3$ GeV higher than a nucleon and therefore $N\Delta$- and $\Delta\Delta$-diagrams produce peaks which are around $0.3$ GeV apart.
Consequently, the theoretical uncertainty of our approach affects mostly the region of the second peak since it is driven by the treatment of 2p2h inclusion to ${\rm Im}\Sigma_\Delta$. This can be observed in all the panels of Fig.~\ref{fig:dsigmadE} where this peak is slightly lower and shifted towards higher energy transfers when comparing the ``2p2h'' to ``2p2h (prev)'' results.
The total 3p3h strength for each considered energy is consistent with the results of Fig.~\ref{fig:total}, and it is almost negligible for $E_\nu=0.5$ GeV and then grows up to $E_\nu\approx1.1$ GeV, when it stabilizes thanks to the implementation of the cutoff in the transferred momentum.  
The 3p3h $d\sigma/d q^0$ distribution is shifted towards high energy transfers where larger phase-space is available and, thus, three particles can be easier produced. This contribution has been calculated and parametrized in such a way that there is no direct way to split it into various isospin channels.

The panels in the middle column in Figs.~\ref{fig:dsigmadE} and \ref{fig:dsigmadTh} show distributions for two protons in the final state. They clearly dominate over the process in which a neutron-proton pair is produced, as shown in the right column. We should point out that NEUT  leads to much higher cross sections for the $np$ pair. In principle in the region of high energy transfers, where $\Delta\Delta$-diagram dominates, the proportion of $pp$ to $np$ should be in the ratio five to one~\cite{Nieves:2011pp,Gran:2013kda}. However, such pattern is not followed by the NEUT sample. The difference, might be partially due to the implementation of the 3p3h contribution within NEUT.

In all the panels of Figs.~\ref{fig:dsigmadE} and \ref{fig:dsigmadTh}  we observe some differences between the NEUT results and the two predictions for the 2p2h. Nevertheless the agreement is sufficiently fair to confirm the reliability of the two-nucleon absorption neutrino cross sections previously published~\cite{Nieves:2011pp} and widely used by the neutrino community. There are three reasons responsible for the existing discrepancies: (i) the contribution of 3p3h is only included into the NEUT calculation (ii) the change of $g'$ to $g'_l(p)$ and $g'_t(p)$, of the $\Delta$ propagator and the $C_A^5$ form-factor, (iii) the abandonment of the use of any averaged nucleon momentum in the computation of the hadron tensor. 
The change of $g'$ from a constant to a momentum-energy dependent function influences predominantly the contribution coming from the $\Delta\Delta$- and $N\Delta$-diagrams  (see the left and the right panels of Fig.~\ref{fig2:2_delta}). Other changes enlisted in (ii) and (iii) do not have a great impact on the final result. In fact, partially they even compensate, since (iii) tends to lower the cross sections, while the new value of  $C_A^5$ produces a certain enhancement. 

\subsection{Available energy}
\label{sec:available}
In several neutrino oscillation experiments the neutrino energy is reconstructed using a calorimetric approach, i.e. by measuring the energy deposited by the outgoing hadrons simultaneously detecting the final lepton.
For this kind of analysis (e.g. performed by the MINERvA experiment~\cite{Rodrigues:2015hik}) the concept of available energy is used, as an attempt to reconstruct the total hadronic energy. For this comparison, we estimate the available energy as the sum of the kinetic energy of all the protons leaving the nucleus. 

The MINERvA results~\cite{Rodrigues:2015hik} point out to a deficit of  events with two nucleons in the final state in the predictions obtained from the implementation  of theoretical model derived in Ref.~\cite{Nieves:2011pp}, within the GENIE Monte Carlo event generator. The 1p1h and 2p2h models used in this Monte Carlo generator are similar to the ones implemented in NEUT. The purpose of this section is to identify whether and how a more accurate description of the final state affects the prediction of the available energy. 

When a neutron-proton pair is produced, the visible energy strongly depends on how the energy is distributed between the two nucleons.
In turn, when two protons are produced, in first approximation, the total energy of the final state should be just a function of the energy transferred  to the final state hadrons, and thus it should not depend on how the energy is shared between the two final protons. This statement might not be totally correct and can be altered by the final state transport of the nucleons through the nucleus. This is because the intranuclear cascade depends on the kinetic energy of the traveling nucleons. It is therefore critical to convolute the predictions of the model for the first step (primary reaction) with a reasonable description of the nucleon transport in the nucleus, accounting for secondary collisions.   To perform this calculation we used migration matrices, generated by the NEUT's cascade, which transform the neutron and proton kinetic energies obtained at the primary vertex into the distribution of energy of outgoing protons after the internuclear cascade. Nucleons suffer mostly elastic scatterings, which cause a smearing of the kinetic energy. Only a small ratio of neutrons transform into protons. Note that these latter effects were not considered in the previous Figs.~\ref{fig:dsigmadE} and \ref{fig:dsigmadTh}.

\begin{figure}[h]
	\centering
	\includegraphics[scale=0.66]{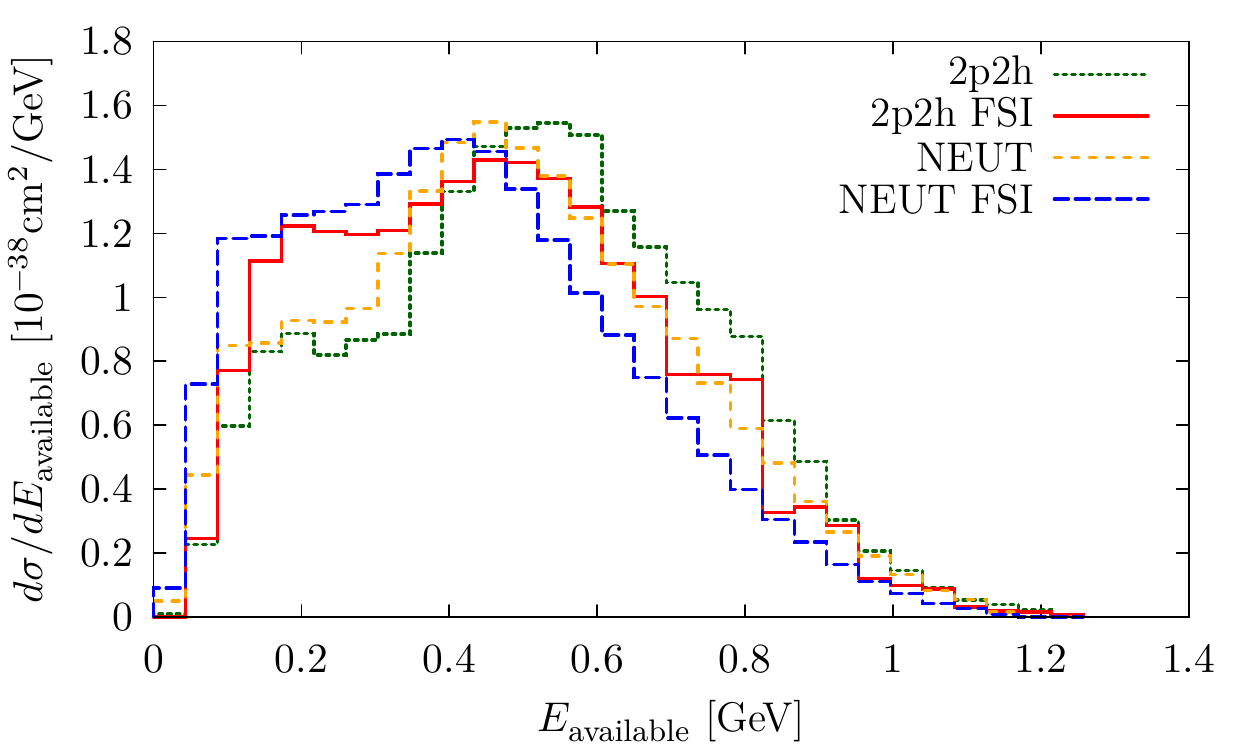}
	\includegraphics[scale=0.66]{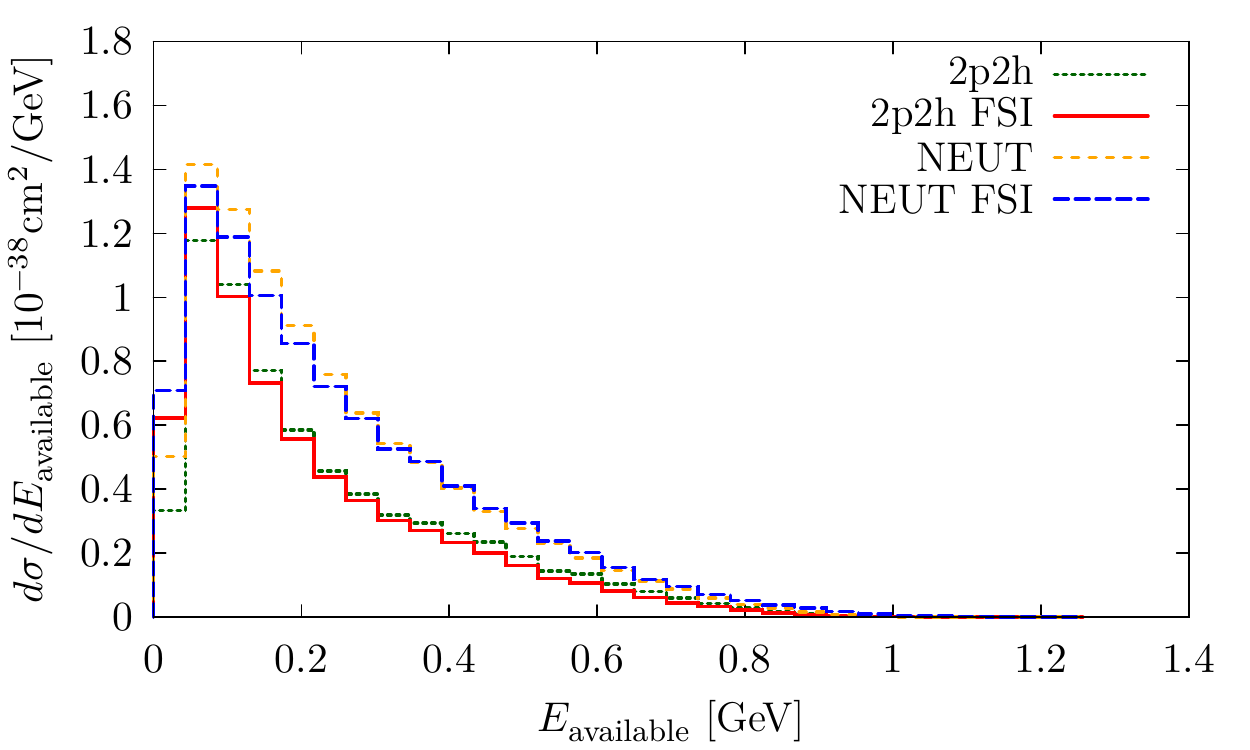}
	\caption{Available energy for $E_\nu=1.5$ GeV. Left: Results for two protons produced in the primary vertex. Right: Results for proton-neutron final state in the primary vertex. Curves denoted as ``FSI'' were obtained with the migration matrices generated with the  NEUT Monte Carlo event generator. 
}
	
	\label{fig:available}
\end{figure}

Using the NEUT migration matrices, we obtain the energy distribution of the outgoing protons after the cascade, or equivalently  the available energy. In Fig.~\ref{fig:available} we show such distribution both for NEUT and for our present calculation,  for $E_\nu= 1.5$ GeV. 
In the left panel we show the situation for two protons in the final state. The difference in normalization comes from the differences between NEUT and present calculation mentioned before. Nevertheless, we observe very similar shapes of the distributions. This should not be surprising for the primary vertex, since -- as we have said before -- the total kinetic energy of the outgoing protons depends on the energy transferred to the nucleus and not on the details of how the energy is distributed between the two outgoing nucleons. The cascade smears the two-peaks structure (whose origin we have already attributed to two dynamical mechanisms represented by the $N\Delta$ and $\Delta\Delta$ diagrams), and shifts strength towards lower energies.

The right panel of Fig.~\ref{fig:available} corresponds to the neutron-proton final state. 
In all the cases the distribution peaks at around $0.075$ GeV (corresponding to a proton with a momentum of approximately 375 MeV). We do not observe the characteristic two-peak structure and the effect of the intranuclear cascade is milder than for the proton-proton final state. 
Still, the difference between the two approaches is well visible. Apart from the normalization factor, the tail of the distribution is much steeper for the ``2p2h'' predictions. This behaviour can be better understood when we look at the two dimensional momentum distributions of the final nucleons, presented in the next subsection.

\subsection{Outgoing nucleons distribution}
\begin{figure}[h]
\centering
\includegraphics[scale=0.42]{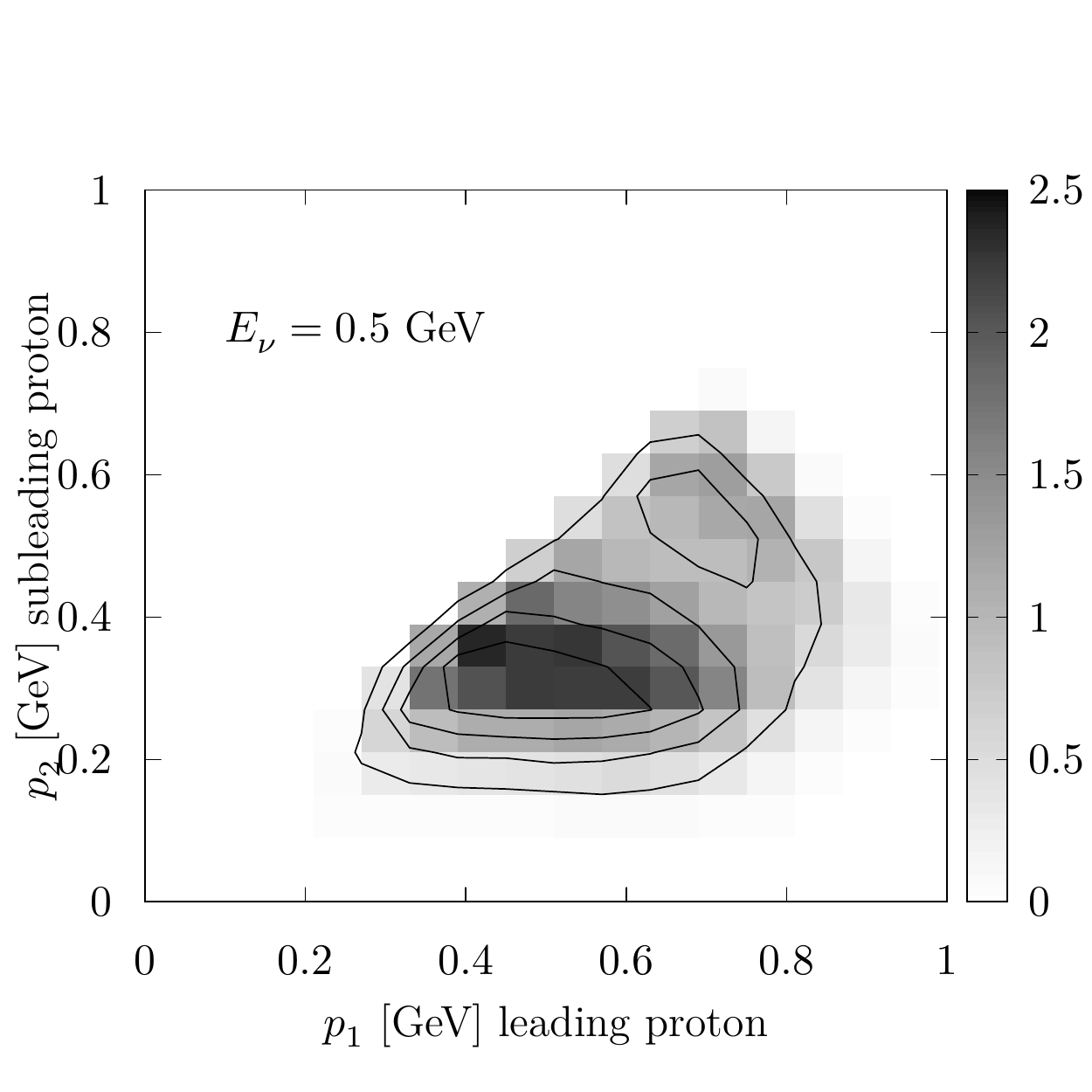}
\includegraphics[scale=0.42]{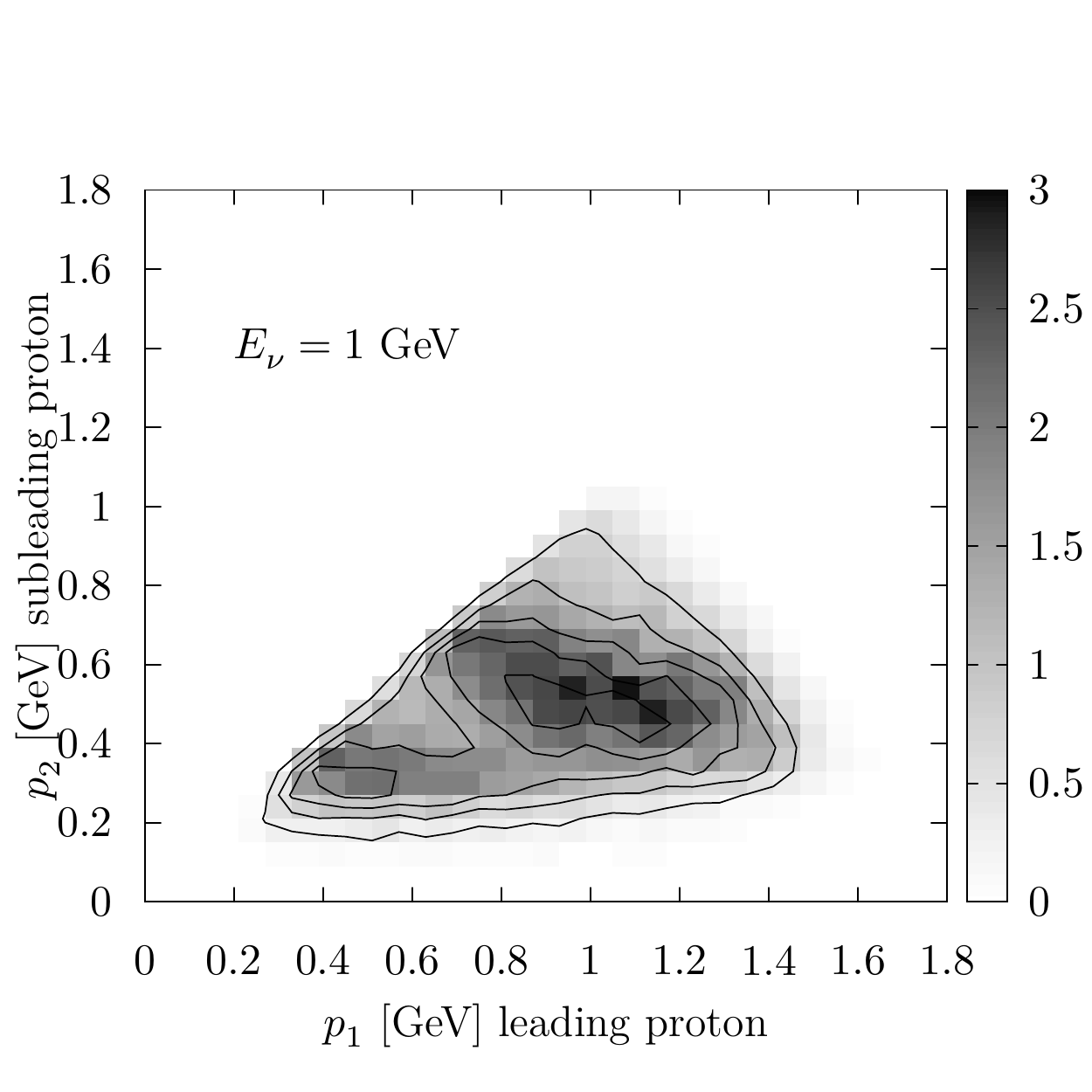}
\includegraphics[scale=0.42]{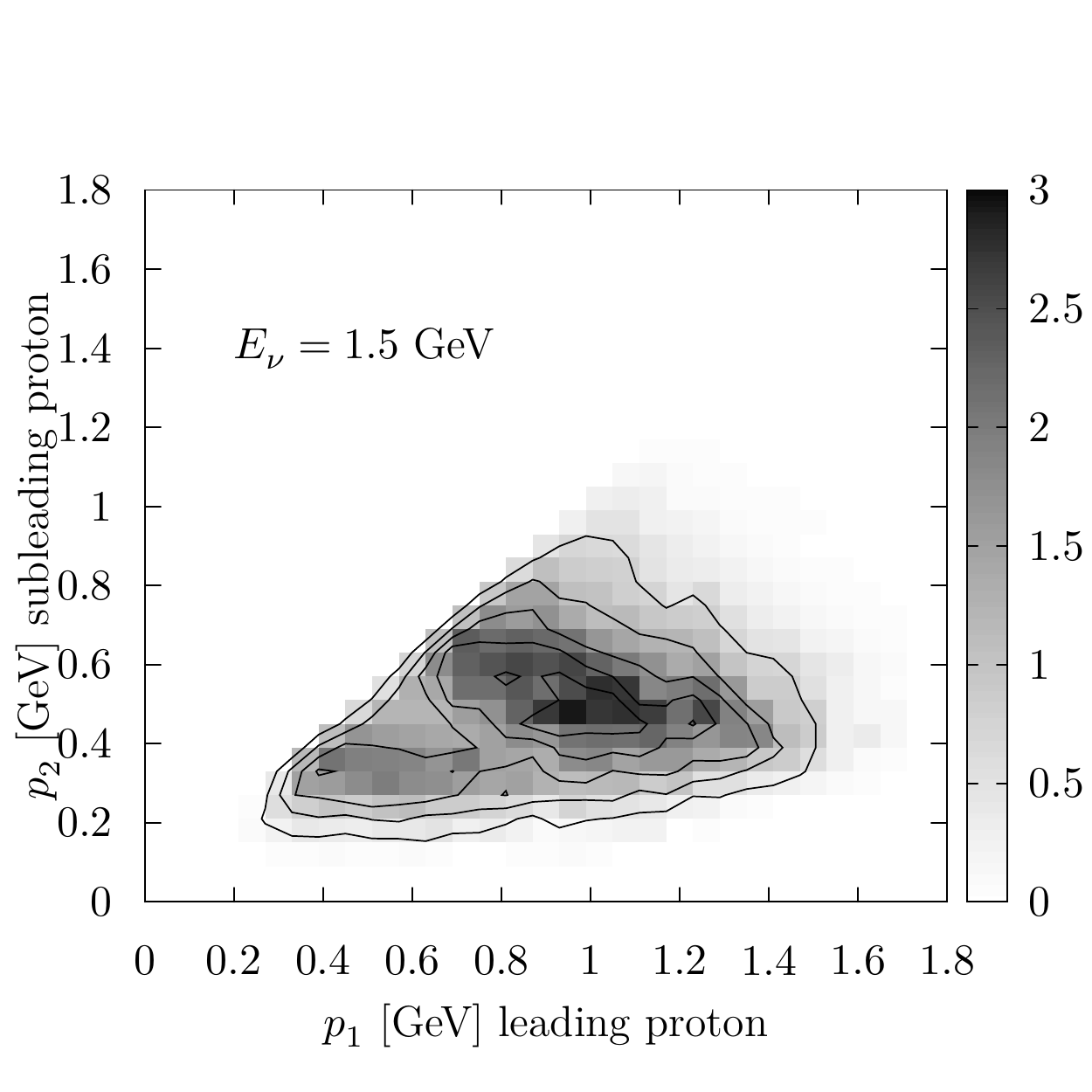}

\includegraphics[scale=0.42]{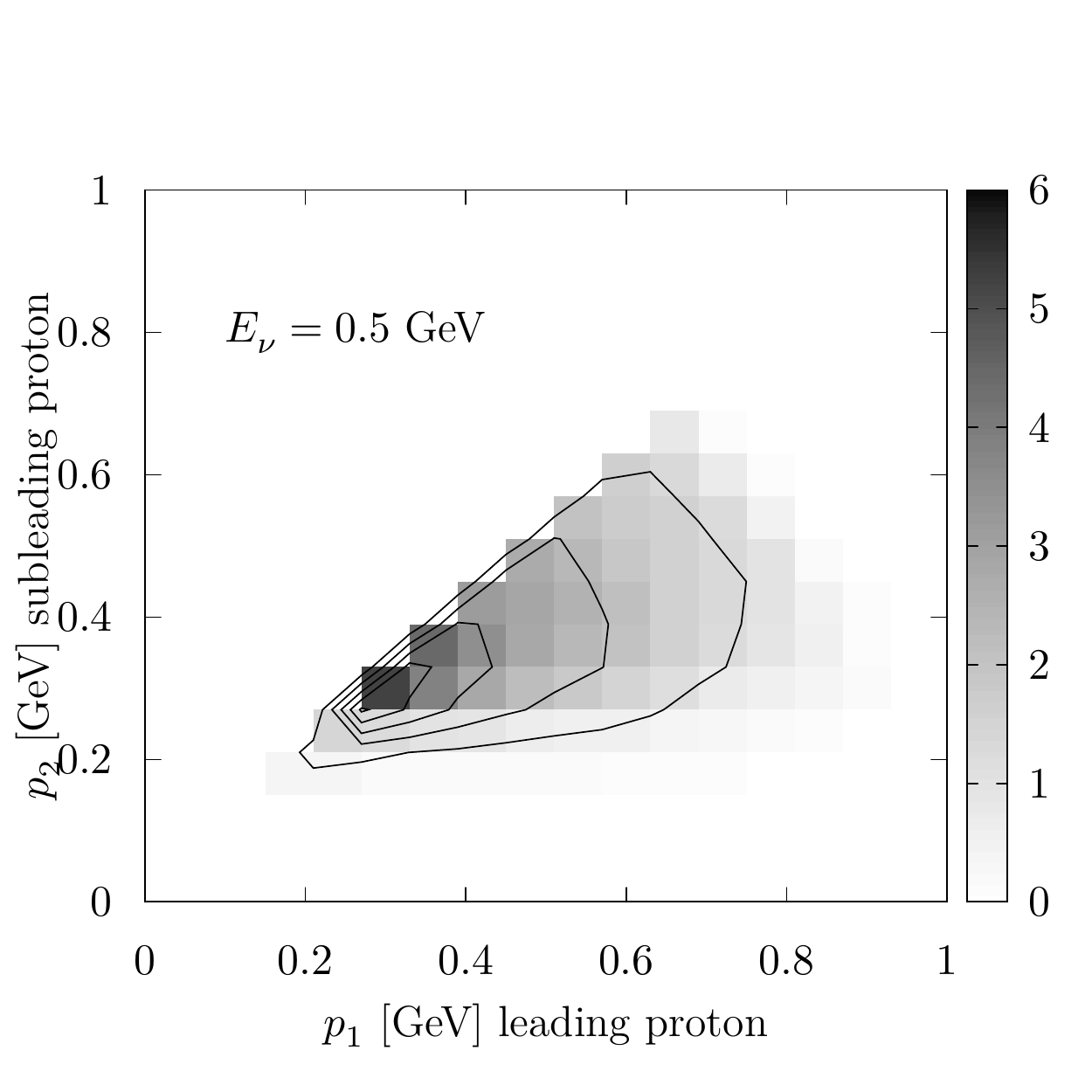}
\includegraphics[scale=0.42]{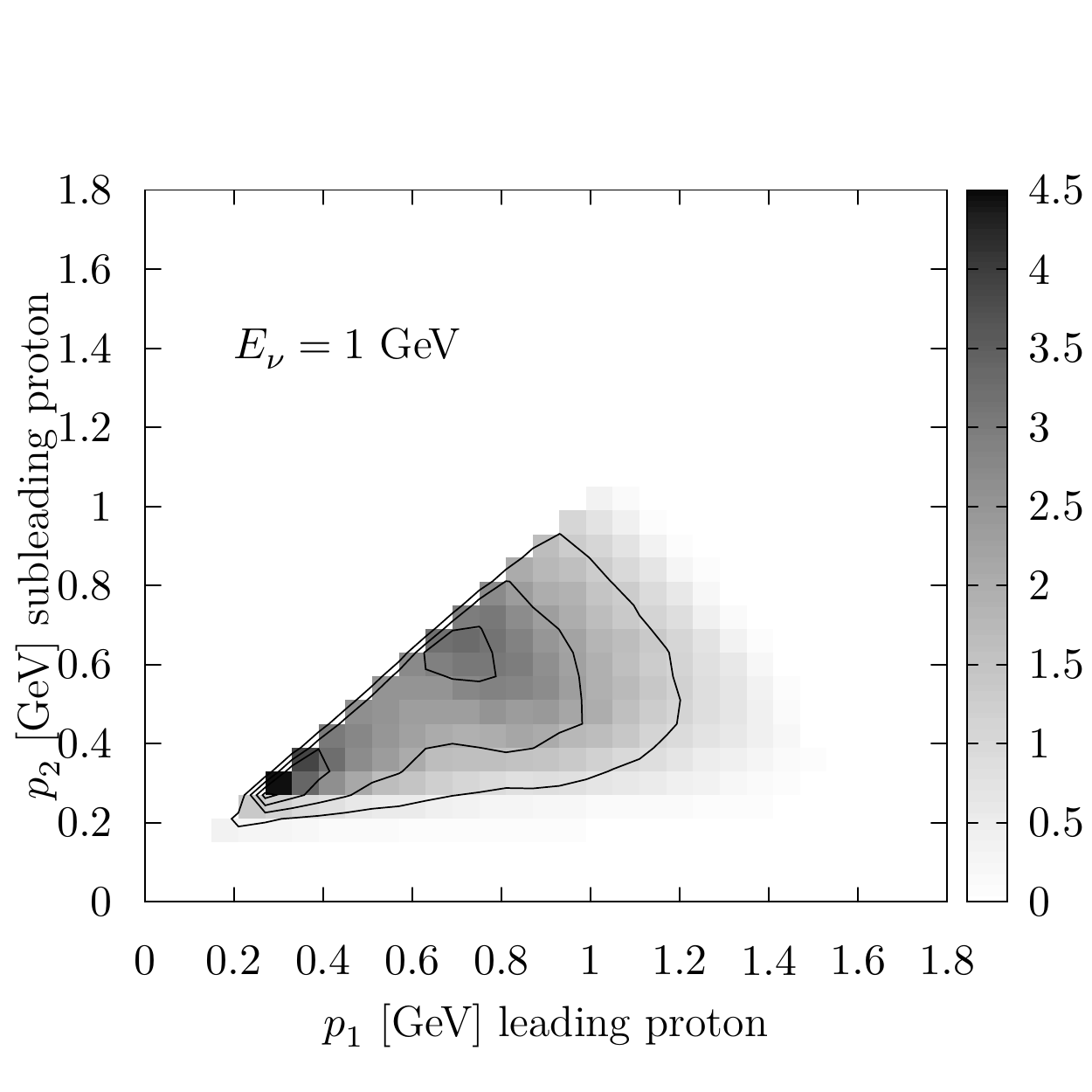}
\includegraphics[scale=0.42]{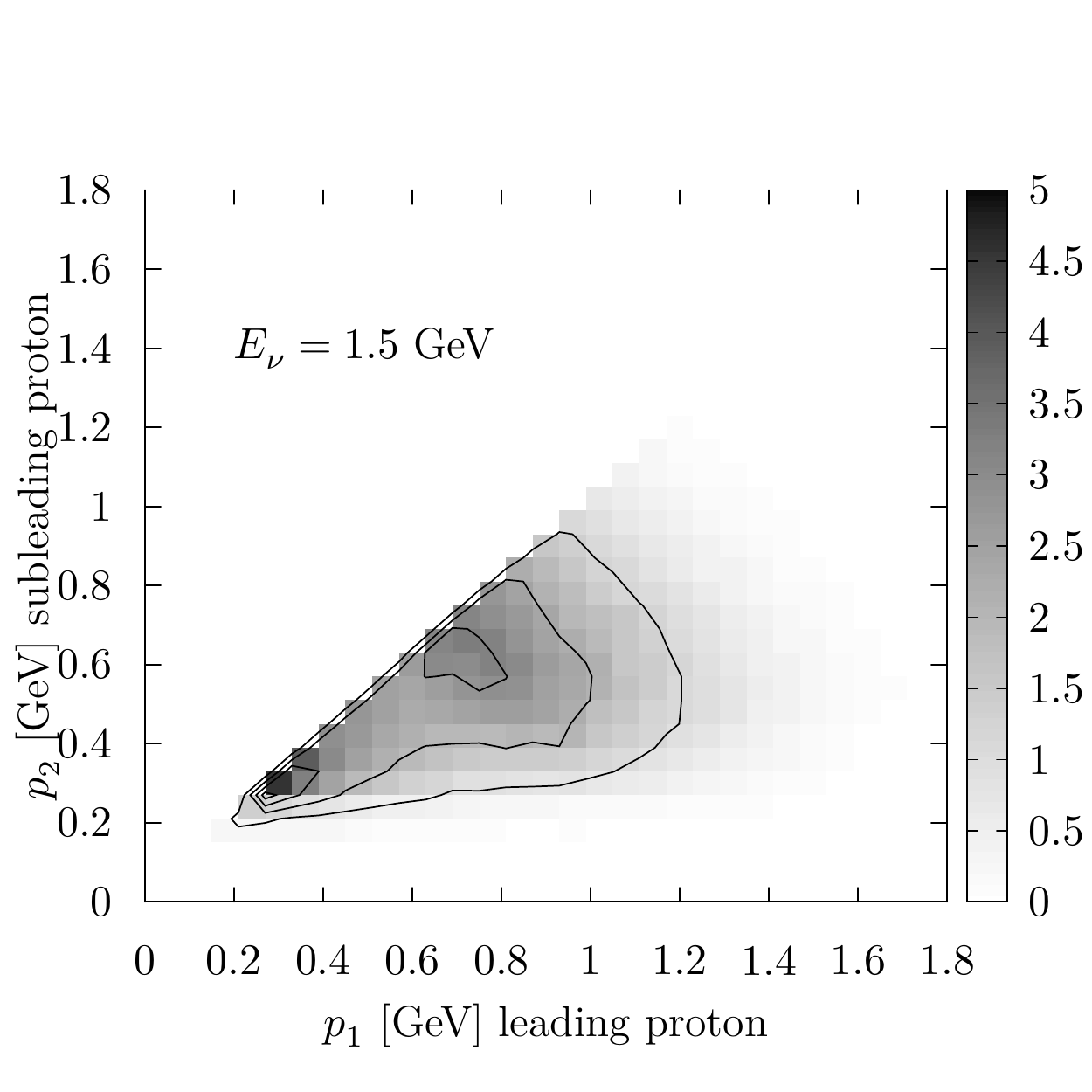}
\caption{Outgoing nucleons distribution $d\sigma / \ d |\vec{p}_1| d|\vec{p}_2|$ $[10^{-38}\text{cm}^2/\text{GeV}^2]$ for two protons in the final state. The momentum $|\vec{p}_1|$ corresponds to higher energetic (leading) proton, while $|\vec{p}_2|$, to the subleading one. The panels from the left to right are for incoming neutrino energies of $E_\nu=0.5$, 1 and 1.5~GeV, respectively. In all cases, the target nucleus is $^{12}$C. Upper panels show the results for the ``2p2h'' model, while the bottom ones have been obtained using NEUT.}
\label{fig:nucleons2D_leading}
\end{figure}

When considering the $d\sigma / \ d |\vec{p}_1| d|\vec{p}_2|$  distribution of two nucleons in the final state, we will analyze separately the proton-proton and neutron-proton cases. In the latter situation the particles are distinguishable, while for two protons, we will look at the distribution of higher- versus lower-energetic proton. This cross section is shown in Fig.~\ref{fig:nucleons2D_leading}.
In the upper panels the results of our current ``2p2h'' calculation are presented. They differ substantially from the distributions provided by NEUT, displayed in the bottom panels. The NEUT results are symmetric (let us point out that the scale used in the bottom panels is around twice larger than for the upper ones).
For $E_\nu=1$, $1.5$ GeV we observe two dominating peaks of the distributions. The first one -- which is also present for the $E_\nu=0.5$ GeV case -- corresponds to the momenta of $p_1\approx0.4-0.5$~GeV,  $p_2\approx0.3-0.4$~GeV (upper panels) and $p_1\simeq p_2\simeq0.3$~GeV (lower panels). This peak is generated by the $N\Delta$-diagram (see the right panel Fig.~\ref{fig2:2_delta}). 
Another, much wider peak is visible for $p_1\approx0.9-1.2$~GeV,  $p_2\approx0.4-0.5$~GeV (``2p2h'' results in the upper panels) and for $p_1\simeq p_2\simeq 0.7$~GeV (NEUT results), generated by the $\Delta\Delta$-diagram. 
It is worth noting that the current ``2p2h'' results predict much higher rate of energetic nucleons in comparison to NEUT. This observation might influence the experimental analysis since the number of observed protons  depends on the detector threshold. It might happen that when the energy is distributed equally between two protons (like it is done in NEUT) they both would lie below the detection threshold and no signal would be predicted. In this case, the asymmetric energy distribution favours the detection of the more energetic one.

\begin{figure}[h]
	\centering
	\includegraphics[scale=0.42]{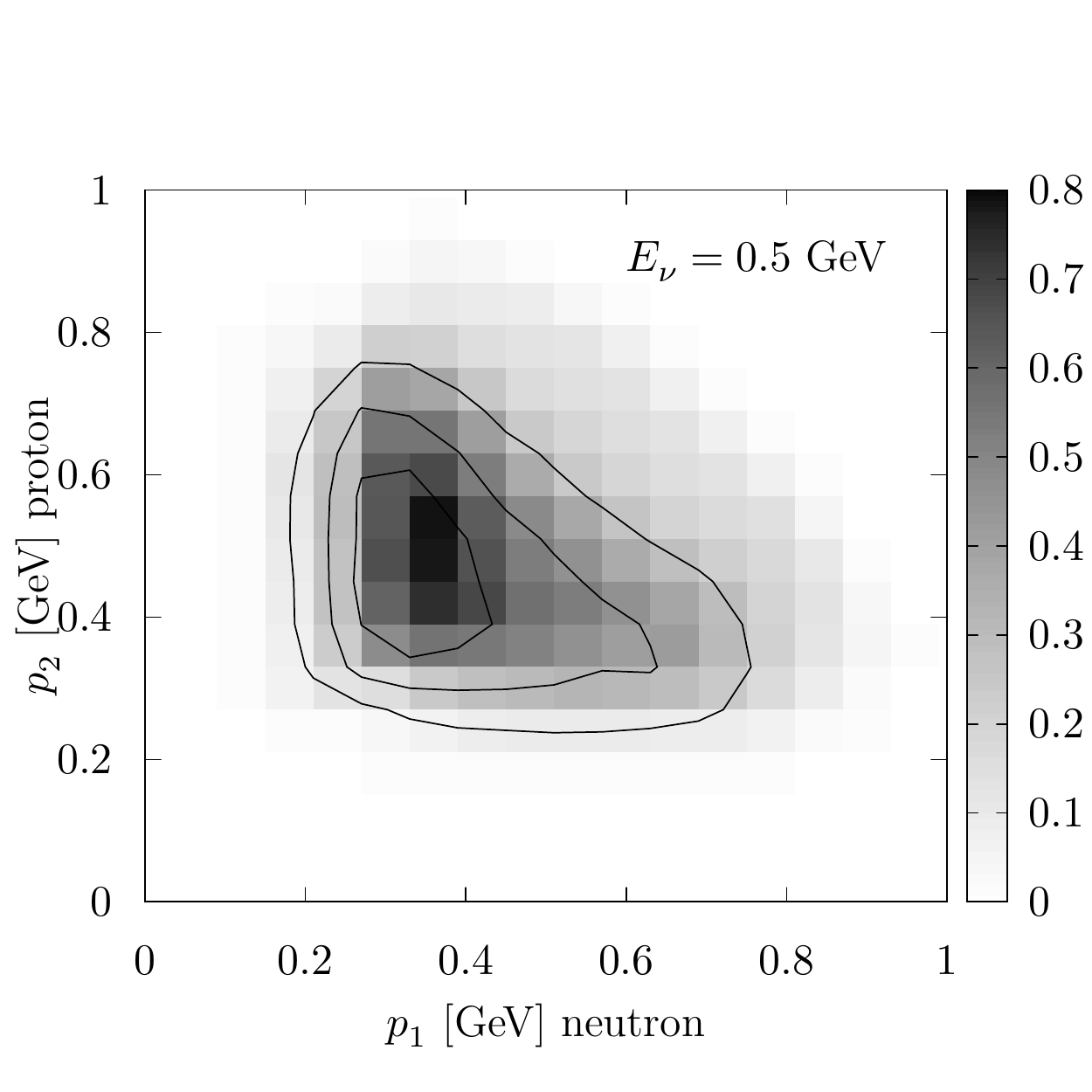}
	\includegraphics[scale=0.42]{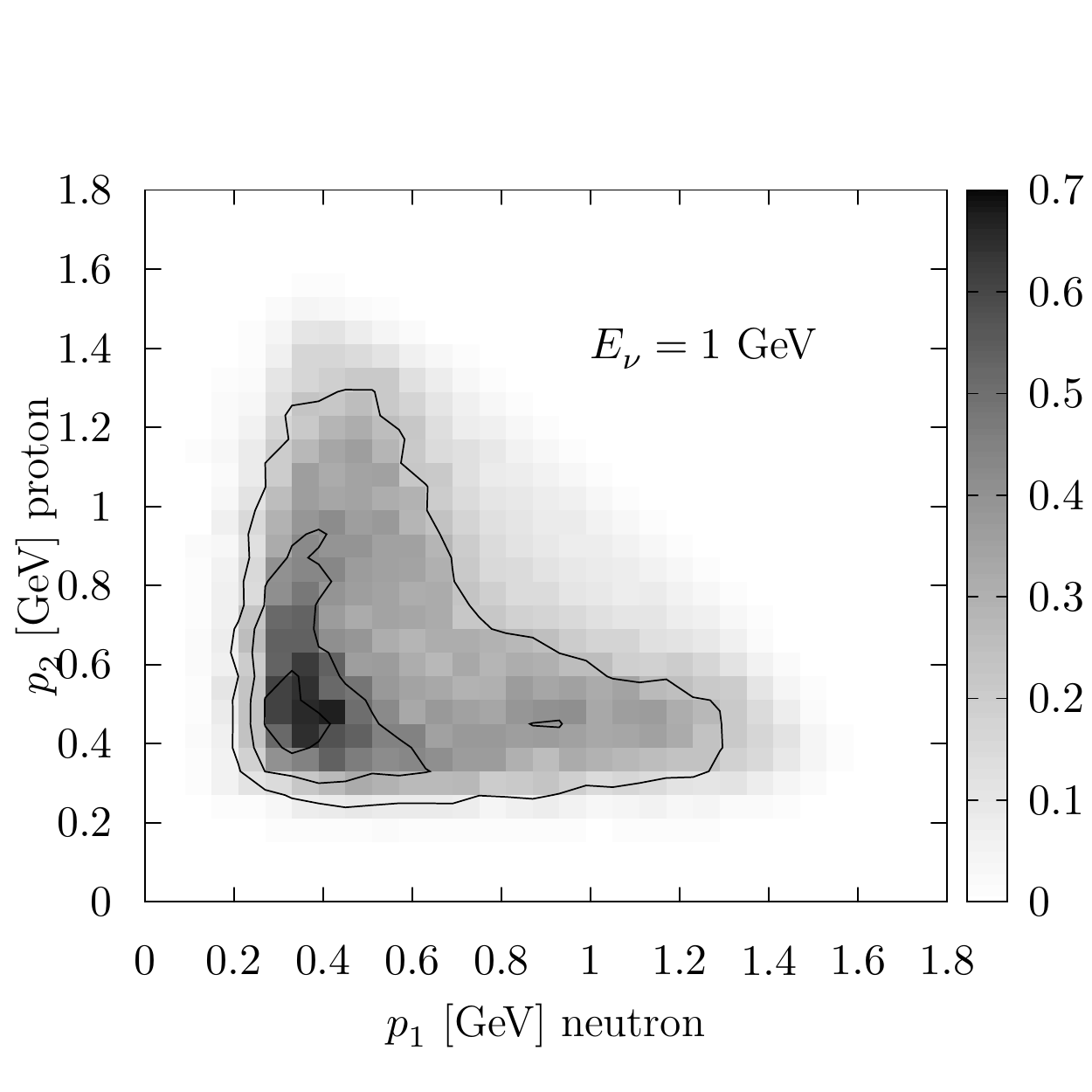}
	\includegraphics[scale=0.42]{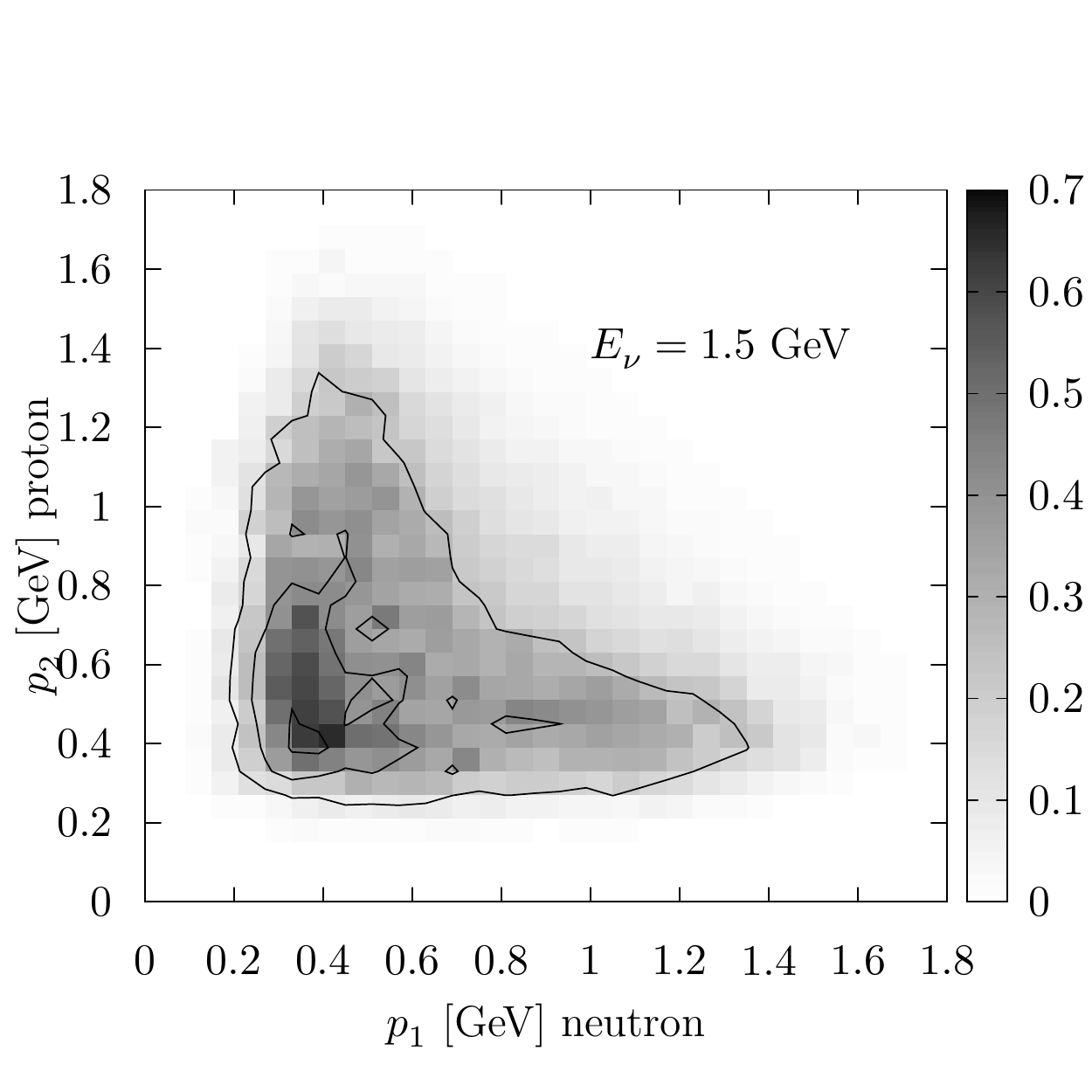}
	
	\includegraphics[scale=0.42]{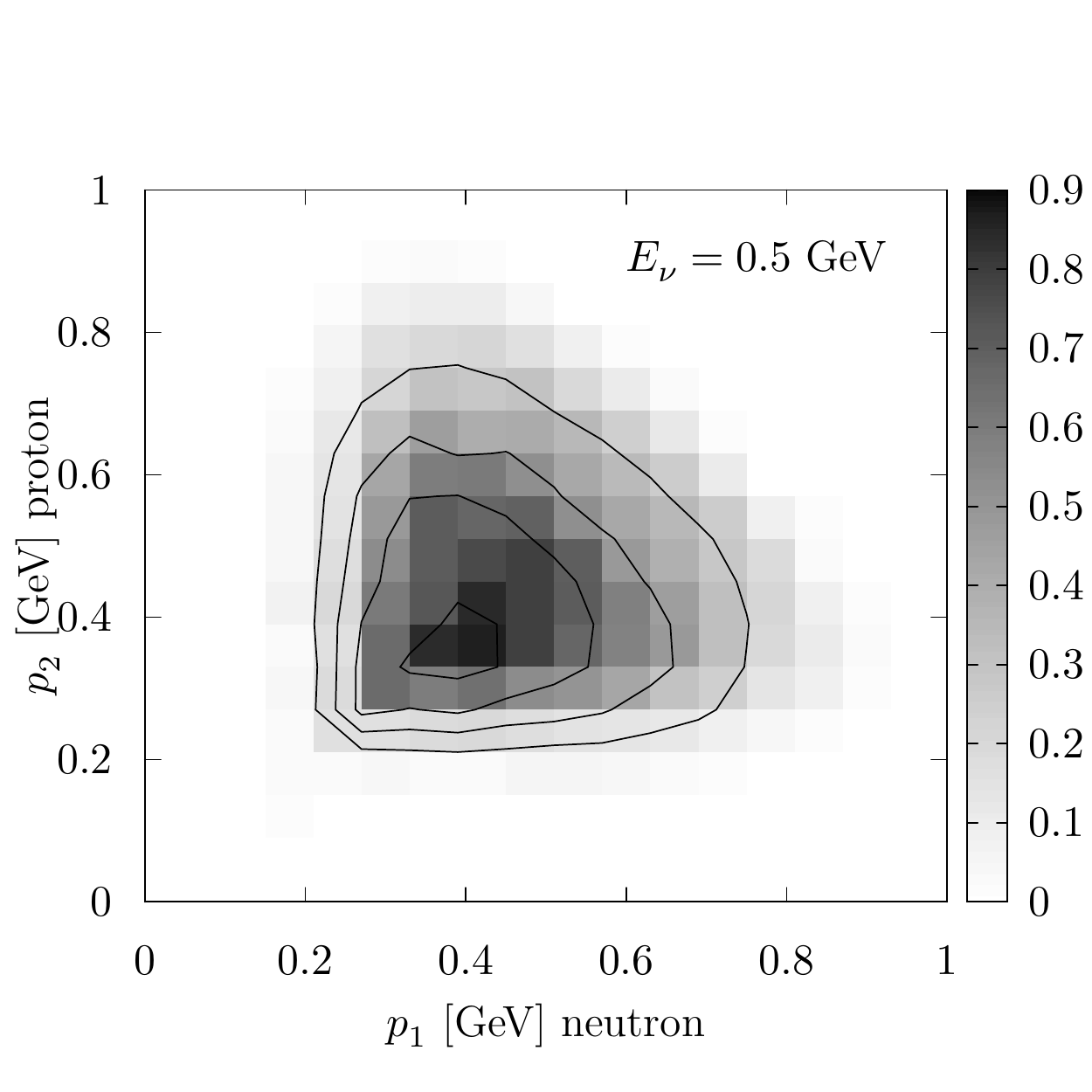}
	\includegraphics[scale=0.42]{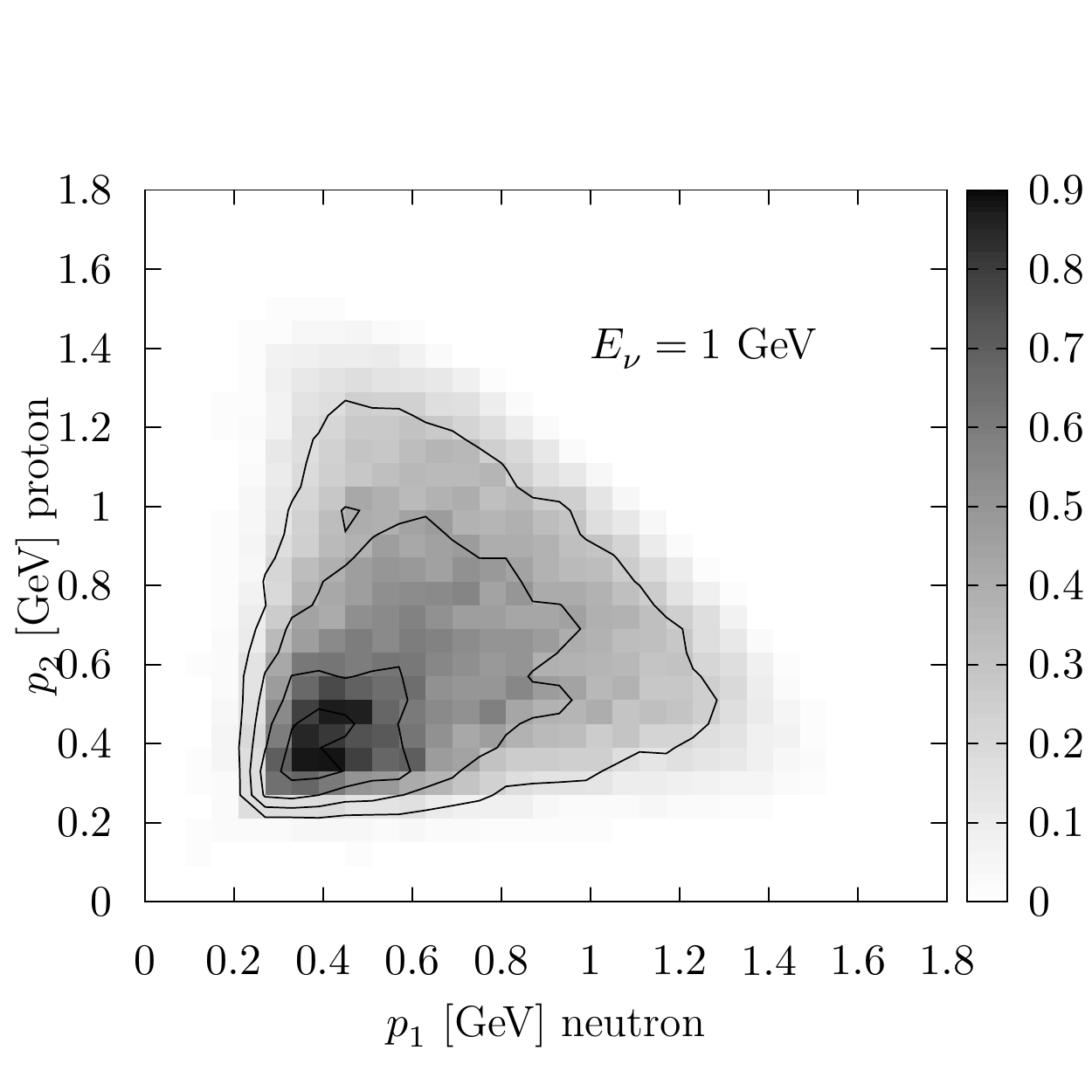}
	\includegraphics[scale=0.42]{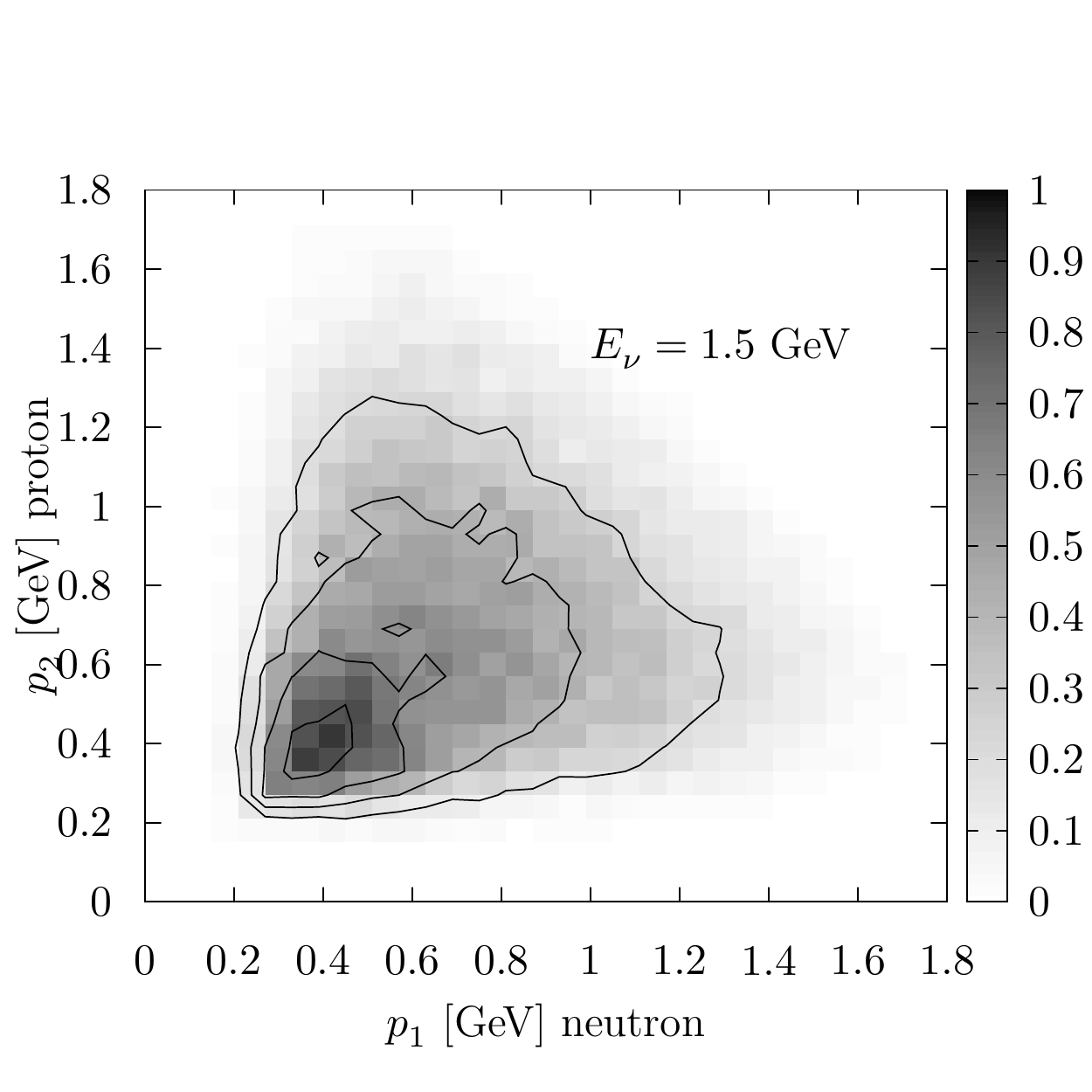}
	
	\caption{Outgoing nucleons distribution $d\sigma / \ d |\vec{p}_1| d|\vec{p}_2|$ $[10^{-38}\text{cm}^2/\text{GeV}^2]$ in the case of neutron-proton pair produced in the final state. The panels from the left to right correspond to incoming neutrino energy $E_\nu=0.5$, 1, 1.5 GeV on $^{12}$C.  Upper panels show the results for the ``2p2h'' model, while the bottom ones for NEUT.
	}
	\label{fig:nucleons2D_np}
\end{figure}

For the case of neutron-proton final state, the cross section is much lower than in the proton-proton case. In Fig.~\ref{fig:nucleons2D_np} we show results for $E_\nu=0.5$, 1 and 1.5 GeV. We observe that the distribution is symmetric for the NEUT (bottom panels), as it follows directly from the procedure of how the nucleons' momenta are generated. Our model predicts a quite different distribution although much more symmetric than in the case of proton-proton production. 
This shape can be understood when we realize that there are two main contributions to the cross section. The interaction on the initial neutron-neutron pair may produce a proton in the ph excitation directly connected to the $W^+$  boson. In this case, a much higher energy-momentum is transferred to the proton. Inversely, when a neutron is produced in the leading ph, it becomes much more energetic.
These two possible contributions have similar strengths (which can be also inferred when analyzing the Clebsh-Gordan coefficients of both channels). When they are summed, the resulting ``L'' shape is obtained, clearly visible in the upper panels of Fig.~\ref{fig:nucleons2D_np}.

%
%
%
%

\section{Conclusions and outlook}
\label{sec:conclusions}
We have presented a revised calculation of 2p2h mechanism induced by CC neutrino scattering, following the theoretical approach of Ref.~\cite{Nieves:2011pp}, and confirmed the reliability of previous results, widely used by the neutrino cross-section community.
In this work we lifted some of the approximations used previously. In particular, the momentum and energy dependence of $g'_l$ and $g'_t$ parameters of the effective in-medium nucleon-nucleon potential has been retained. This allowed us to calculate directly the inclusion of the $\Delta\Delta$-diagrams which beforehand were added using a parametrized result from Ref.~\cite{Oset:1987re}.
Also, we have updated the way how the $\Delta$ propagator is treated, retaining only the spin $3/2$ part using consistent couplings, and we have included the most recent value of the $C_A^5$ form-factor fitted to the available electroweak pion production ANL and BNL data. 
The difference between our current and previous calculations can be treated as a theoretical error of our approach. An overall uncertainty in the total cross section does not exceed $10\%$. 

Although the distributions of the outgoing leptons are not much affected by the introduced changes, the experimental analyses usually depend also on other observables and details of theoretical models. 
This motivated our further inquiries into the hadronic final states. With the NEUT migration matrices we were able combine our model predictions in the primary vertex of interaction with realistic estimates of  the internuclear cascade effects.

We also separated the 2p2h and 3p3h contributions and consequently -- with the above mentioned updates -- we gained an insight into the distribution of outgoing nucleons, considering different isospin channels.
We have compared the outgoing nucleons' momentum distributions, coming directly from our microscopic approach,  with the ones obtained within NEUT, where the nucleons generated isotropically in the hadronic center of mass according to the available phase-space. The procedure followed by NEUT is implemented in other Monte Carlo event generators, so the results of our analysis can be directly applied also to them.

In the case of proton-proton final state, we predict a strong asymmetric signal with one of the protons more energetic than another. For the neutron-proton case, the obtained distribution is also quite different from the isotropic result.
How these predictions would enter and alter the existing experimental analyses depends on the used detection techniques. Certainly, some differences should be visible due to the proton's momentum thresholds. Also, it has been often argued that the information about exclusive hadronic states in the neutrino-nucleus scattering is an important observable which can help to discern between various theoretical approaches.

According to our predictions, the 3p3h contribution is important, and might even amount to $20\%$ of the 2p2h strength for energy-momentum transfer regions.
It would be therefore interesting and useful to redo the calculation of Ref.~\cite{Oset:1987re}, in the same way as we did for the 2p2h, to get further insight into the strength coming from different isospin channels of this process. The transferred energy this time would be divided into three outgoing nucleons, thus making them on average less energetic (and more difficult to observe because of the detector's threshold).
We suspect that -- the same way as in the 2p2h case -- the momentum distribution of the outgoing particles would be asymmetric with one leading nucleon.

Further analysis of multinucleon knockout cross sections for antineutrino-induced and neutral-current driven processes are  natural continuation of the present work and we are already working on them. 
Available antineutrino-nucleus scattering data is less accurate due to lower event-rate (and therefore higher statistical uncertainties), leading to weaker constraints for theoretical models. Still, this channel will play a crucial role in the experimental programs aiming at measuring the CP-violating phase.
The NC 2p2h is a less explored channel due to the experimental difficulty of performing a measurement. However,  it should also be included in the Monte Carlo generators, treating it as a background process. An analysis of how its presence affects the experimental studies is hence a important  topic for future work.

\begin{acknowledgements}

This work was supported by the Swiss National Foundation Grant No. 200021\_85012, the Spanish Ministerio de Economia y Competitividad and the European Regional Development Fund under contract FIS2017-84038-C2-1-P, the EU STRONG-
2020 project under the program H2020-INFRAIA-2018-
1, grant agreement no. 824093, the Cluster of Excellence “Precision Physics, Fundamental Interactions, and Structure of Matter (PRISMA$^+$),” funded by the German Research Foundation (DFG) within the German Excellence Strategy (Project ID 39083149), and by the DFG-funded Collaborative Research Center SFB 1044.
\end{acknowledgements}

\bibliography{biblio}

\end{document}